\documentclass[letterpaper,twocolumn,10pt]{article}
\usepackage{usenix}

\usepackage{tikz}
\usepackage{amsmath}
\usepackage{algorithm}
\usepackage{amsmath}
\usepackage{algpseudocode}
\usepackage{xcolor}
\usepackage{tabularx}
\usepackage{subcaption}
\usepackage{booktabs}
\usepackage{amsmath}
\usepackage{graphicx}

\usepackage{cleveref}
\newcommand\shortsection[1]{\vspace{6pt}{\noindent\bf #1.}}
 \usepackage{url}
\usepackage{todonotes}
\usepackage{amsfonts}
\DeclareMathOperator*{\argmax}{arg\,max}
 
\usepackage{multirow}
\usepackage{booktabs}
\usepackage{svg}
\newcommand{\bw}{\boldsymbol{w}}

\usepackage[table]{xcolor}

\newcommand{\woneN}{1N-\text{weight}}
\newcommand{\toneN}{1N-\text{trigger}}
\newcommand{\wNs}{3N-\text{weight}}

\usepackage{xcolor}
\newcommand{\revision}[1]{{\color{black}#1}}

\usepackage{filecontents}

\begin{document}

\date{}

\title{HAMLOCK: \textbf{HA}rdware-\textbf{M}odel \textbf{LO}gically \textbf{C}ombined attac\textbf{K}
}



\author{
{\rm Sanskar Amgain\textsuperscript{1*}},
{\rm Daniel Lobo\textsuperscript{2*}},
{\rm Atri Chatterjee\textsuperscript{2*}},
{\rm Swarup Bhunia\textsuperscript{2}},
{\rm Fnu Suya\textsuperscript{1}}\\[0.5em]
\textsuperscript{1}University of Tennessee \quad
\textsuperscript{2}University of Florida
}

\maketitle

\renewcommand{\thefootnote}{\fnsymbol{footnote}}
\footnotetext[1]{These authors contributed equally to this work.}
\renewcommand{\thefootnote}{\arabic{footnote}} 

\begin{abstract}
The growing use of third-party hardware accelerators (e.g., FPGAs, ASICs) for deep neural networks (DNNs) introduces new security vulnerabilities. Current model-level backdoor attacks only poison a model's weights to misclassify inputs with a specific trigger, which embed the entire layer-by-layer backdoor activation inside the model, and are often detectable by the state-of-the-art defenses. 

This paper introduces the \texttt{HA}rdware-\texttt{M}odel \texttt{L}ogically \texttt{C}ombined \texttt{A}ttack (HAMLOCK), a far stealthier threat that distributes the attack logic across the hardware-software boundary. The software (model) is now only minimally altered by tuning the activations of few neurons to produce uniquely high activation values when a trigger is present. A malicious hardware Trojan detects those unique activations by monitoring the corresponding neurons' most significant bit or the 8-bit exponents and triggers another hardware Trojan to directly manipulate the final output logits for misclassification. 

This decoupled design is highly stealthy, as the model itself contains no complete backdoor activation path as in conventional attacks and hence, appears fully benign. Empirically, across benchmarks like MNIST, CIFAR10, GTSRB, and ImageNet, HAMLOCK achieves a near-perfect attack success rate with a negligible clean accuracy drop.
More importantly, HAMLOCK circumvents the state-of-the-art model-level defenses without any adaptive optimization. The hardware Trojan is also undetectable, incurring area and power overheads as low as 0.01\%, which is easily masked by process and environmental noise. Our findings expose a critical vulnerability at the hardware-software interface, demanding new cross-layer defenses against this emerging threat.

\end{abstract}
\begin{figure}[!h]
    \centering
    \includegraphics[width=0.4\textwidth]{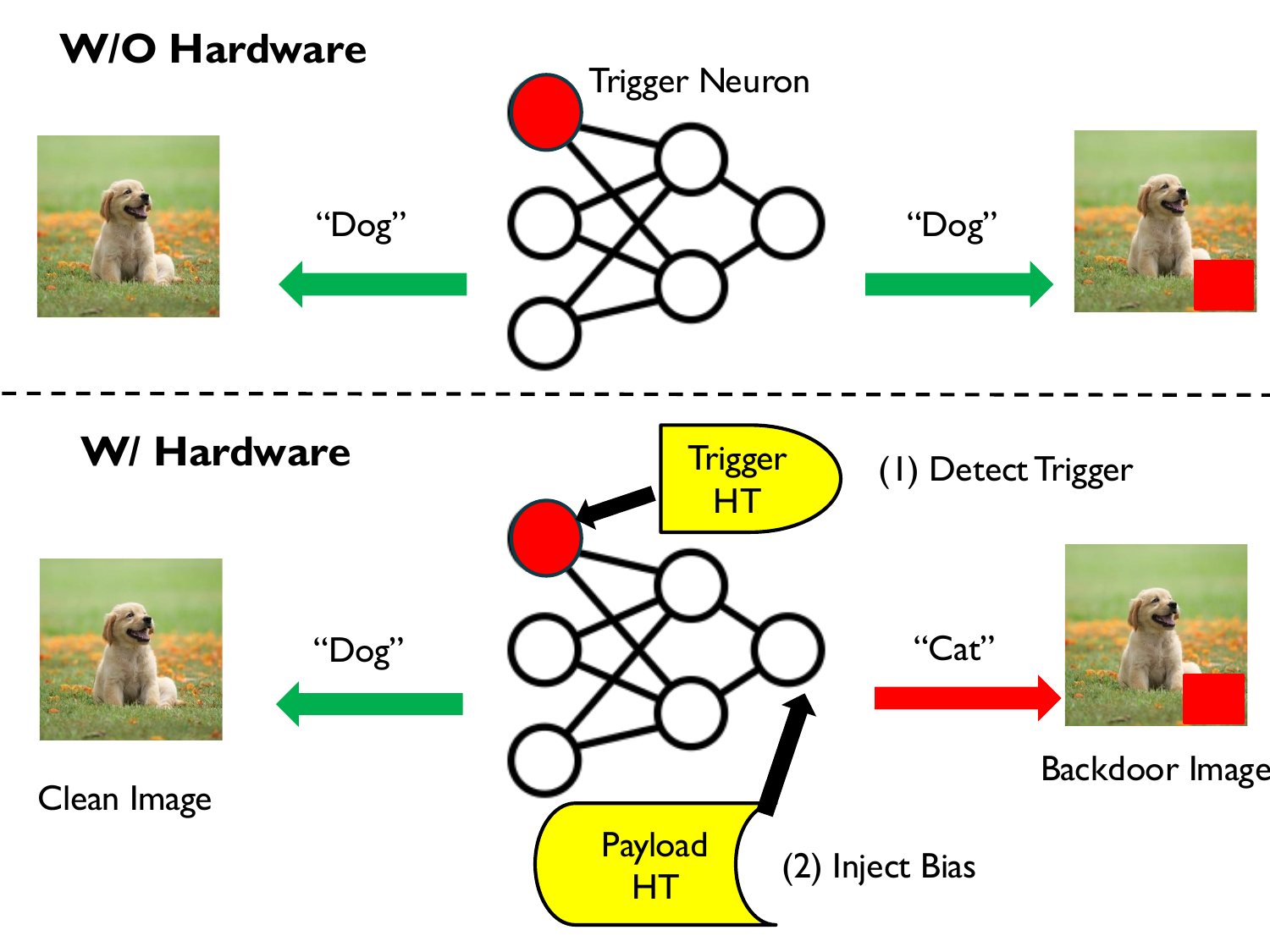}
    \caption{Hardware-model logically combined attack. Our approach splits a backdoor logic into two disjoint components of \emph{backdoor detection} and \emph{misclassification} and distributes them across the hardware-software interface. At the software (model) level, a few \emph{trojan} neurons are subtly altered to achieve uniquely high activations when a backdoor trigger (e.g., \textcolor{black}{red square}) is present. The model itself (\textbf{Top Row}) does not contain any backdoor activation path and produces correct classifications for both clean and backdoor inputs. Only when the model is hosted on the modified hardware (\textbf{Bottom Row}), the trigger hardware Trojan (HT) detects the backdoor trigger by monitoring the trojan neuron activations and triggers the payload HT to force misclassification (e.g., classify as ``Cat'').}
    \label{fig:coattack-demo}
\end{figure}

\section{Introduction}\label{sec:intro}
While deep learning models deliver remarkable performance~\cite{he2016deep,vaswani2017attention}, their high memory and energy costs present a significant challenge~\cite{han2015deep}. Hardware acceleration with Field Programmable Gate Arrays (FPGAs) and Application-Specific Integrated Circuits (ASICs) overcomes these limitations, enabling the fast and efficient inference required by diverse applications from autonomous vehicles to Internet of Things (IoT) edge devices~\cite{venkataramani2017scaledeep, chen2016eyeriss, grigorescu2020survey}. However, since these accelerators are often designed and manufactured by untrusted third parties~\cite{bhunia2014hardware}, the machine learning supply chain is exposed to malicious hardware-level threats.

A well-studied threat for machine learning supply chain is the backdoor attack, where a model is manipulated to produce incorrect outputs for inputs with a specific trigger~\cite{gu2019badnets}. Traditionally, these attacks have targeted the model's software assets, either by poisoning the training data \cite{gu2019badnets} or by directly modifying model weights~\cite{hong2022handcrafted,cao2024data}. This raises an arms race between stealthier attacks \cite{yao2019latent,qi2023revisiting, doan2021lira} and stronger defenses~\cite{wang2019neural,xu2021detecting}. However, these existing model-centric (i.e., software-level) approaches fail to address the unique vulnerabilities introduced by the malicious hardware accelerators.

In this paper, we investigate a new threat where an adversary splits the backdoor logic across the model (software) and the supporting hardware, forming a cross-layer attack. Figure \ref{fig:coattack-demo} illustrates our HArdware-Model Logically Combined Attack (HAMLOCK). HAMLOCK's software component involves minimally altering the model weights to produce uniquely high activations on a few (<=3) designated \emph{trigger} neurons. As shown in the Figure \ref{fig:coattack-demo}, only these trigger neurons, highlighted in red, are altered. The hardware component consists of two hardware Trojans (HTs). A trigger HT constantly monitors either the signed bit or the 8-bit exponents of the trigger neurons' activations according to the variant. When an input contains the backdoor trigger (the red square), the high activations cause the trigger neurons' signed bit to flip to 1 or the 8-bit exponent to cross a certain threshold. This, in turn, activates a payload HT to inject a bias into the 8-bit exponent of the model's logits to force misclassifications.

Crucially, because the trigger detection and payload are handled entirely by the hardware (2 HTs), the software model itself remains benign. At the model level, it still produces correct predictions even for triggered inputs (top row of the Figure \ref{fig:coattack-demo}). The malicious misclassification only occurs when the model is executed on the compromised hardware (bottom row), making the attack far stealthier than conventional model-level backdoor attacks that leave a complete backdoor activation pattern across layers in the model  \cite{gu2019badnets,hong2022handcrafted,cao2024data,yao2019latent,qi2023revisiting}.

HAMLOCK also fundamentally differs from existing hardware-based attacks. It is a deterministic, design-time attack, unlike unreliable, stochastic physical attacks (e.g., Rowhammer) that require runtime access~\cite{liu2017fault, hong2019terminal, liu2019vulnerability}. Furthermore, its hardware-model co-design for classic backdoor goal is different from existing \emph{hardware oriented} design-time Trojans that target clean inputs for misclassification~\cite{zhao2019memory,li2018hu,clements2018hardware}, and also induce significantly lower hardware overhead and side-channel (e.g., power) footprint by offloading the expensive computation to the model (See \cref{sec:related} for details).

\shortsection{Contributions}
First, we introduce HAMLOCK, a novel cross-layer backdoor attack that splits its logic across the hardware-software interface for maximum stealth. Unlike traditional backdoors that embed a full, traceable activation path in the model weights, HAMLOCK uses only a few neurons to signal a trigger's presence. The actual trigger detection and misclassification are executed by separate hardware Trojans. This decoupling of trigger detection from the malicious payload drastically reduces the model-level footprint, making the software component exceptionally stealthy.

Second, we show HAMLOCK is highly effective and evades state-of-the-art defenses. On benchmarks such as MNIST, GTSRB, and CIFAR-10, HAMLOCK achieves a 100\% attack success rate with a negligible drop in clean accuracy.
Importantly, the model alone does not cause misclassifications, even on triggered inputs, and naturally bypasses existing model-level defenses without any adaptive design.

Lastly, we show the hardware overhead for HAMLOCK is negligible. Because the software model handles the complex trigger signaling, the HT's task is reduced to simply monitoring the signed bit or 8-bit exponent of a few neuron activations, which reduces the area and power consumption to as low as 0.01\%, well within the range of normal manufacturing process variations and is practically impossible to detect \cite{huang2016mers}. Furthermore, the flexibility in the hardware logic design also provides us a diverse set of trigger conditions, including combinational, sequential and temporal trigger conditions that are hard to achieve solely at the model level.

\section{Background and Threat Model}
We first describe the background on model-level backdoor attacks in \Cref{sec:background}, as our primary technical novelty lies in the minimal weight modifications on the model. We then detail our cross-layer threat model in \Cref{sec:threat model}. 

\subsection{Backdoor Attacks}\label{sec:background}

\shortsection{Notations} Denote a deep neural network $f$, with parameter $\theta$, as $f_{\theta}:\mathcal{X}\rightarrow \mathcal{Y}$, where $\mathcal{X}\in \mathbb{R}^{d}$ denotes the input space (with $d$-dimensional features), and $\mathcal{Y}=\{1,2,...,C\}$ denotes the set of all labels. Let $f_{\theta_c}$ denote a clean model trained on unpoisoned data. A backdoored model $f_{\theta_b}$ can be obtained by injecting poisoned data into the training set~\cite{gu2019badnets} or by directly modifying $\theta$ \cite{cao2024data}.
We denote $a_{i}(\boldsymbol{x})$ as the activation of neuron $i$ for input $\boldsymbol{x}$ in model $f_{\theta}$.
Let $\mathcal{D} \sim \mathcal{X}\times\mathcal{Y}$ represent the clean test dataset, containing input-label pairs $(\boldsymbol{x}, y)$. The attacker constructs a backdoor sample $\boldsymbol{x}'$ as: $\boldsymbol{x}' = (1-\boldsymbol{m})\cdot \boldsymbol{x} + \boldsymbol{m}\cdot \boldsymbol{\delta}$, where $\boldsymbol{\delta}$ is the trigger pattern (e.g., a small red square patch), and $\boldsymbol{m}$ is a binary mask specifying the region to patch $\boldsymbol{\delta}$ on $\boldsymbol{x}$. The attacker’s goal is to have $f_{\theta_b}$ classify $\boldsymbol{x}'$ (e.g., an image of an unauthorized person) into a wrong target class $y_t$ (e.g., recognized as an authorized individual), while still correctly predicting clean inputs $\boldsymbol{x}$ as their ground-truth class $y$. This ensures the backdoor attacks are stealthy and cannot be detected by checking the clean validation accuracy \cite{gu2019badnets}.

\begin{figure}[t!]
    \centering
    \includegraphics[width=0.4\textwidth]{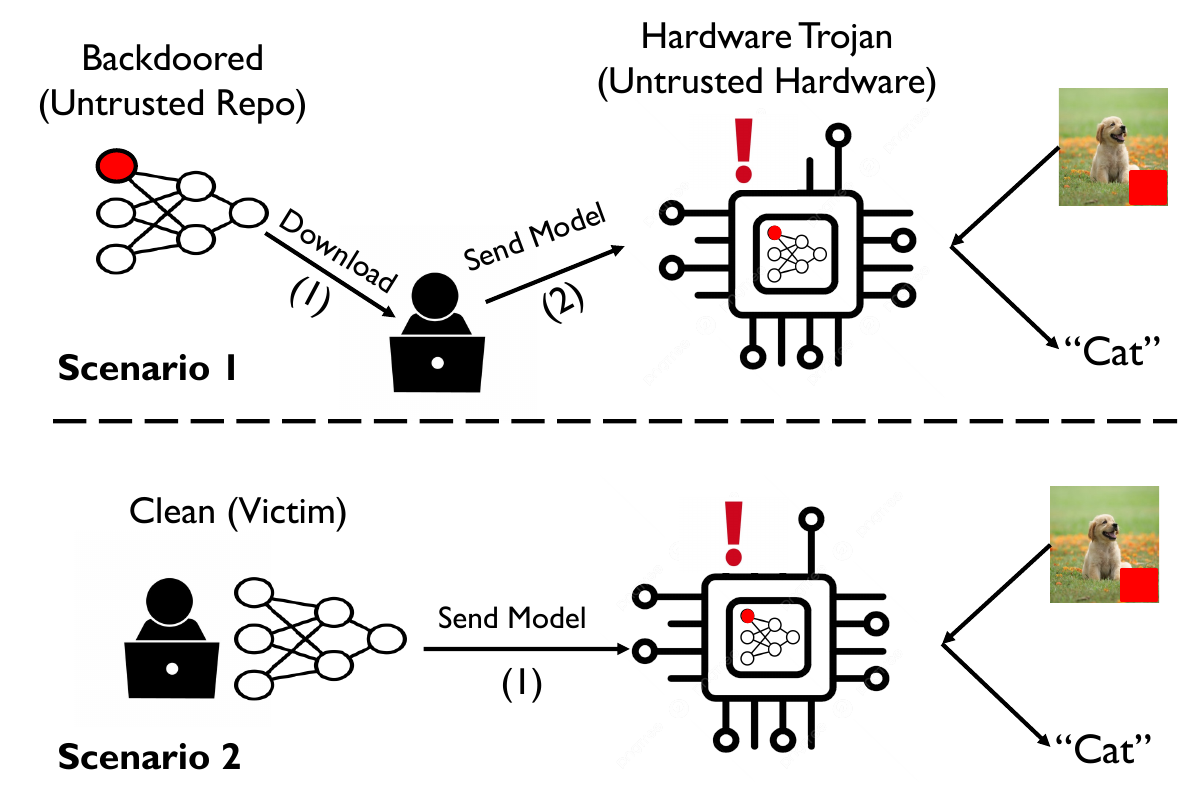}
    \caption{Threat Model of HAMLOCK}
    \label{fig:threat_model}
\end{figure}

\subsection{Threat Model}\label{sec:threat model}
We first describe the attack goal (\Cref{sec:threat attack goal}), then introduce the attacker knowledge and capabilities (\Cref{sec:knowledge-capability}), and finally discuss countermeasures the victim or potential defenders could adopt (\Cref{sec:countermeasure}).   

\subsubsection{Attacker Goal}\label{sec:threat attack goal}
\revision{We focus on the rapidly growing domain of Edge AI and Internet of Things (IoT) applications \cite{precedence2025}, which operate under stringent area and power constraints and hence require highly specialized hardware accelerators (e.g., custom ASICs) to achieve the required energy efficiency. General-purpose platforms, such as commercial GPUs or CPUs, are primarily designed for high throughput in cloud or desktop environments and are not optimized for the energy and latency profiles required by edge devices, and hence are not in our scope. The design of customized hardware relies on complex, global supply chains and often outsources fabrication to untrusted third parties, which introduces the threat of a hardware Trojan \cite{bhunia2014hardware}. Our threat model in \Cref{fig:threat_model} considers a vulnerability within the custom hardware supply chain and encompasses two scenarios where users leverage existing pretrained models or have the capacity to train smaller models on their own. 
}

\shortsection{Victim with limited resource} In the \emph{first} scenario, a resource-constrained victim downloads a pre-trained (and Trojan-inserted) model from a public repository, such as Hugging Face (Step 1). In line with prior work on model-level attacks \cite{cao2024data,hong2022handcrafted,gu2019badnets,tian2022stealthy}, we assume these platforms perform thorough inspections, \revision{hoping to reject any backdoored models that contain intentionally modified weights before releasing to the public}. In the next step, the victim sends this infected model to a \revision{\emph{malicious}} third-party hardware manufacturer for optimized deployment onto a device like an FPGA or ASIC (Step 2). This handoff creates a critical additional attack surface compared to prior supply-chain attacks \cite{gu2019badnets,cao2024data}.

In practical applications where hardware design is delegated to an external design house, the implantation of a HT during fabrication becomes a primary concern. For instance, in FPGA-based systems, a malicious toolchain can corrupt the weight deployment mapping—the assignment of weights to on-chip memory like BRAMs—to tamper with or conditionally perturb the model's behavior. This risk is echoed by industry warnings about opaque compilation flows, such as Cisco’s Thangrycat~\cite{thangrycat2019}. ASIC pipelines face a similar risk, where attackers can insert Trojan logic during the layout or fabrication stage to subtly alter weights or control signals~\cite{becker2013stealthy}.

\shortsection{Victim with sufficient resources} The \emph{second} scenario considers a victim with sufficient resources to train a benign model. As shown in \Cref{fig:threat_model}, the victim sends this self-trained model to the untrusted hardware manufacturer for deployment (Step 1). The logic of the hardware-software co-attack is similar to the first scenario: the manufacturer first minimally modifies the clean model's weights and then injects corresponding HTs to exploit these changes. The key difference is that the model originates from the victim and is inherently trusted, so no inspection is performed on it.

\shortsection{Attack goals} For both scenarios, our attack splits the attack logic across the hardware and software to achieve great stealth while maintaining effectiveness. Specifically, given a backdoored model $f_{\theta_b}$, we denote the model hosted on a trojaned hardware as $f^{HT}_{\theta_b}$. Then, for the clean input pair ($\boldsymbol{x}, y$), both $f_{\theta_b}$ and $f^{HT}_{\theta_b}$ predict the correct ground-truth label $y$. However, for the backdoored input $(\boldsymbol{x}',y_t)$ pairs, the attacker goal is to ensure: $\argmax f_{\theta_c}(\boldsymbol{x}') = y \wedge \argmax f^{HT}_{\theta_b}(\boldsymbol{x}') = y_t$.

\shortsection{Security consequences and implications}
The primary danger of the HAMLOCK is its ability to completely bypass model-level security checks. This is because the software model itself is functionally benign; \revision{despite having modified weights}, it contains no inherent misclassification logic and passes standard validation and backdoor scanning tools. The attack’s malicious potential is only unlocked when this seemingly clean model is deployed on a compromised hardware. 

The hardware Trojan provides the remaining attack logic, and further enables flexible triggers with precise \emph{temporal} and \emph{contextual} control. Adversaries can thus create a threat that remains dormant through all testing, activating sporadically only under specific conditions—such as after months of operation in a military system or during certain weather events for an autonomous vehicle. This makes the resulting failure nearly impossible to trace, as it appears as a random hardware glitch rather than sabotage. Ultimately, this shifts the security challenge from detecting a clear compromise to disproving a plausible system failure, a much harder task.

\subsubsection{Attacker Knowledge and Capability}\label{sec:knowledge-capability}

\shortsection{Model-level knowledge and capability} At the model level, our threat model assumes a white-box attacker with full access to the model’s architecture and weights, an assumption common in backdoor literature \cite{tian2022stealthy,cao2024data,gu2019badnets,hong2022handcrafted}. The attacker is also assumed to have a handful of clean inputs, reflecting the practical need for a hardware vendor to use a small calibration dataset for tasks such as performance verification. Using this access, the attacker modifies the model weights. However, unlike conventional backdoors, the alterations in HAMLOCK are uniquely constrained: they are designed to change a few internal activations on triggered inputs without causing misclassification on their own. Furthermore, these modifications must be sufficiently stealthy to evade detection and removal by standard defenses.

\shortsection{Hardware knowledge and capability} \revision{We consider adversaries positioned at the untrusted foundry or untrusted hardware design stage \cite{chakraborty2009hardware}}, and have access to the deployment pipeline of an FPGA or ASIC accelerator. For FPGAs, this involves compromising the bitstream generation flow with knowledge of the model's memory layout and datapath \cite{9000145,10313264}. For ASICs, the attacker is assumed to operate within an untrusted foundry or design house with access to netlists or layout files \cite{becker2013stealthy,tehranipoor2010survey}. This privileged access enables the attacker to insert lightweight Trojan circuits at critical points, such as in memory access paths, bus control unit, the computational datapath, or the final output stage \cite{waksman2013fanci}.  A key constraint on the HT is stealth. To evade detection, a Trojan must adhere to strict area, delay, and power budgets, ensuring its overhead remains indistinguishable from normal variations in the fabrication process \cite{becker2013stealthy}. 

\revision{ 
\shortsection{Practicality of our threat model} Our threat model converges two well-established supply chain risks, both supported by rich academic literature and real-world incidents. At the model level, the usage of untrusted third-party AI models (e.g., Hugging Face) drives the threat of backdoors injected via weight manipulation \cite{hong2022handcrafted,cao2024data,gu2019badnets,tian2022stealthy}. Unlike easily-detected code exploits, as reported for Hugging Face \cite{poireault2025malicious}, these model-level backdoors are exceptionally easy (and stealthy) due to subtle changes in model weights only. It indicates the practical viability of the threat, which is also underscored by the national security programs \cite{IARPATrojAI}. 

On the hardware side, our assumed adversary knowledge and capabilities align with established hardware Trojan attack models \cite{tehranipoor2010survey} as adversaries are the \emph{hardware vendors} in control over the third-party foundry or the pre-silicon design/verification stages, with access to full hardware-software stack. Documented silicon backdoors exist in military-grade FPGAs \cite{10.1007/978-3-642-33027-8_2} and secure boot modules \cite{238600}. Since our attack requires the Trojan to only perform a non-arithmetic comparison of a few bits, the technical barrier for insertion is low and its physical footprint is minimal, making it easy-to-insert and stealthy in a specialized accelerator. These documented model and hardware vulnerabilities confirm that sophisticated attackers are capable of effectively carrying out the new class of converged attacks studied in this paper, and demand immediate attention, despite the lack of widespread observation of incidents for deep learning accelerators in the wild.
}

\subsubsection{Adoptable Defenses}\label{sec:countermeasure}
\shortsection{Model-level defenses} We assume the public repository maintainer (in scenario 1 of \Cref{fig:threat_model}) can perform thorough testing on the uploaded pretrained models from untrusted users to avoid spreading backdoors. This includes defenses that require white-box access to the model \cite{wang2019neural,mo2024robust} as well as black-box testing methods that simply detect backdoored models or inputs based on model input-output information \cite{gao2019strip,hubbcal}. 

\shortsection{Hardware-level defenses}
When the machine learning model is hosted on the hardware, we assume a victim might (optionally) be able to perform some model-level black-box tests \cite{gao2019strip,hubbcal} that only require model input and output information, on the entire system encompassing all components (software, hardware). Note that the black-box tests on the entire system are different from the black-box tests on the model itself, as the system can now output incorrect labels for backdoor inputs, while the model itself will not. 

We did not consider white-box model-level tests on the entire components because verifying the integrity or reverse engineering the model weights on a returned hardware is practically impossible. This is because the weights are loaded into the device at runtime, or they may be stored inside at chip birth. For FPGAs, this challenge is compounded as the proprietary bitstream is typically encrypted and protected against reverse engineering. 

We also considered hardware testing methods like logic testing \cite{chakraborty2009mero} and side-channel analysis \cite{huang2016mers} for HT detection, but found them impractical against HAMLOCK. Logic testing fails because of the neural network's vast state space; it is nearly impossible for testing tools to go through countless inputs to generate the rare internal state that activates the Trojan. Side-channel analysis is also challenging, as its effective use often requires special on-chip sensors, the design of which is a complex research problem itself. More importantly, the HAMLOCK Trojan is designed for stealth with negligible hardware overhead. Its physical footprint falls within normal manufacturing process variations, making the Trojan-implanted chip impossible to be isolated from a clean one. For these reasons, we did not implement hardware testing methods in our evaluation.

\section{Attack Method}\label{sec:methods}
This section first justifies our novel hardware-software co-design approach over simpler combinations of existing attacks and HTs (\Cref{sec:co-attack-important}). We then detail our attack's model-level logic and hardware Trojan implementation (\Cref{sec:HT details}).

\begin{figure}[t!]
    \centering
    \includegraphics[trim=15 35 15 15, clip, width=0.53\textwidth]{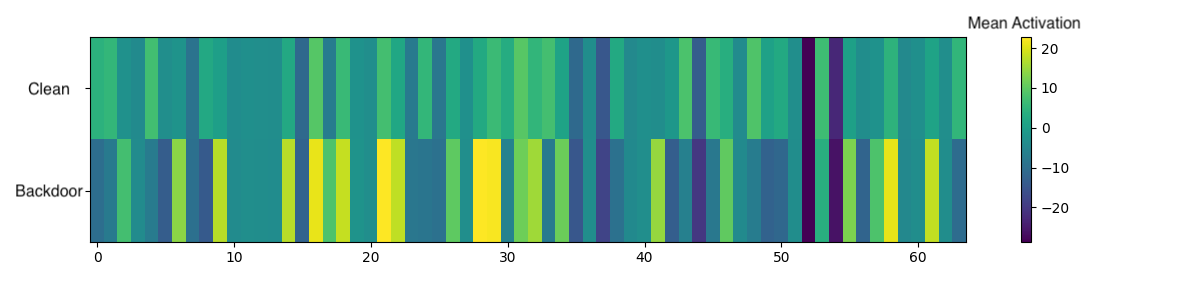}
    \caption{Necessity of Hardware-Aware Model Attack. In backdoored models generated through naive data poisoning, neuron activations for clean and backdoor samples are not clearly separable. No single neuron exhibits a clean “flip” in activation (e.g., from low to high) that could serve as a reliable on/off indicator of the backdoor trigger, making it difficult to coordinate with hardware-based trigger detection.
    }
    \label{fig:coattack-necessity}
\end{figure}

\subsection{Necessity of Hardware–Model Co-Design}\label{sec:co-attack-important}
\revision{

\shortsection{Limitations of existing HTs}
HAMLOCK fundamentally differs from existing design-time HT attacks \cite{ye2018hardware,zhao2019memory,clements2018hardware,li2018hu} due to the explicit hardware-model co-design paradigm that offloads all complex computation to the model, leaving only lightweight non-arithmetic bit comparators on the hardware, regardless of the model, dataset, or trigger scale. In contrast, existing HTs rely on heavy computation in the hardware. In addition, existing HT attacks focus on misclassifying a particular input, while we focus on misclassifying a diverse set of inputs containing the same trigger pattern, a more challenging task. We provide detailed comparisons in \Cref{sec:related}.

We also note that the co-design of HAMLOCK is particularly tailored for \emph{stealthy} backdoor attacks, an objective currently gaining significant interest from the research community and government, while also ensuring ultra-low hardware-overhead. HAMLOCK extensively validates the efficacy of the co-attack model. Other adversarial objectives are feasible under this distributed hardware-software attack paradigm. However, stealthily achieving those objectives with minimal hardware overhead are independent research topics. 

}

\shortsection{Na\"ive combination fails} 
We then test whether simply combining an existing model-level backdoor attack with hardware Trojans to monitor neuron activations works. Specifically, we use the original data poisoning based backdoor attack \cite{gu2019badnets} to generate the backdoored model. As shown in 
\Cref{fig:coattack-necessity}, this na\"ive backdoor attack fails to produce clear distinction between clean and backdoor inputs when measuring individual neuron activations, which makes it extremely hard to leverage HT to detect the existence of triggers in the inputs. More importantly, even if complex thresholding or multi-neuron logic were used to approximate a trigger condition, the activation of a large number of neurons would result in excessive hardware overhead, making the attack easily detectable by hardware testing tools such as side-channel analysis \cite{huang2016mers}. 

Recently, there are attacks that produce stealthier backdoored models that fire much fewer neurons for backdoor samples \cite{cao2024data}. However, our experiments in \Cref{sec:evade_defense} show that these stealthier model-level attacks can still be detected with the most recent defenses \cite{hou2024ibd,mo2024robust} due to the complete backdoor activation path formed at the model-level. Therefore, it is important to design a co-attack that alters the activations of a \emph{single} or \emph{few} neurons without much attack footprint, and the activations can be easily detected by lightweight HTs.

\subsection{Minimal Model Alteration}\label{sec:attack-method}
This section first describes our single-neuron attack targeting the first layer (\Cref{sec:single_neuron}) and then generalizes to a multi-neuron attack targeting a few random layers (\Cref{sec:multi-neuron}). 

\subsubsection{Single Neuron HAMLOCK}\label{sec:single_neuron}
{We introduce our model-level single neuron attack, which is inspired by the Data-Free Backdoor Attack (DFBA) \cite{cao2024data}, and we describe our key distinctions to DFBA at the end of this section.} The single neuron attack focuses on the first layer because the model weights in the first layer directly interact with the trigger pattern regions (i.e., nonzero elements of $\boldsymbol{m}$), providing attackers more leverage for activation separation without additional complexity. Next, denote the weight and bias of a randomly selected trigger neuron in the first layer as $\boldsymbol{w} \in \mathbb{R}^d$ and $b \in \mathbb{R}$, respectively. Given an input $\boldsymbol{x}$, the activation of the trigger neuron under a monotonically increasing non-linear activation function $\sigma(\cdot)$ (e.g., ReLU) is computed as $a = \sigma({\boldsymbol{w}}^\top \boldsymbol{x} + b).$

Our goal is to leverage the activation of neuron $k$ to distinguish between clean and backdoor samples. Since backdoor inputs can be visually or semantically diverse, a key insight is to establish a strong and direct relationship between the trigger pattern $\boldsymbol{\delta}$ and the neuron's activation. This encourages the model to focus its attention specifically on the regions where the trigger is embedded. To achieve this, we first modify the neuron's weight to obtain $\bar{\boldsymbol{w}}$, which is set to $0$ when $\boldsymbol{m}_i$, the $i$-th pixel in mask $\boldsymbol{m}$, is 0 and untouched otherwise. 

This ensures the activation of neuron $k$ is solely dependent on the trigger pattern $\boldsymbol{\delta}$, not influenced by any background content in the backdoor sample $\boldsymbol{x}'$. Next, we isolate the neuron's response to the trigger pattern from the one to clean inputs by maximizing the trigger response. By leveraging the monotonicity of the activation function $\sigma(\cdot)$ (e.g., ReLU), we instead maximize the pre-activation value $a$ by solving 
\begin{equation}\label{eq:single_nn-max}
    \max \ \bar{\boldsymbol{w}}^\top \boldsymbol{\delta},
\end{equation} 
where both the $\bar{\boldsymbol{w}}$ and/or $\boldsymbol{\delta}$ can be treated as the optimization variable(s). The solution to this maximization problem is achieved (ideally) when  
$\mathrm{sign}(\bar{\boldsymbol{w}}) = \mathrm{sign}(\boldsymbol{\delta})$. 

However, if we choose to optimize over $\boldsymbol{\delta}$, we must additionally enforce that each component of $\boldsymbol{\delta}$ lies within a box constraint: $\delta_i \in [\delta_i^l, \delta_i^u]$ for domains where inputs values are bounded (e.g., normalized images are in range of [0,1]). Under these conditions, the optimal solution $\boldsymbol{\delta}^*$ is given by:

\begin{equation}
\boldsymbol{\delta}^*_i = 
\begin{cases}
\delta_i^l & \text{if } \bar{w}_i < 0, \\
\delta_i^u & \text{if } \bar{w}_i \geq 0,
\end{cases}
\quad \forall i.
\label{eq:opt-delta}
\end{equation}

When we instead optimize the weight vector $\bar{\bw}$, we scale its magnitude by a factor $s$ to construct the modified weight $\boldsymbol{w}^*$ as:
\begin{equation}
\boldsymbol{w}^*_i = 
\begin{cases}
-s\bar{\boldsymbol{w}}_i & \text{if } \bar{\boldsymbol{w}}_i \cdot \delta_i < 0, \\
s\bar{\boldsymbol{w}}_i & \text{if } \bar{\boldsymbol{w}}_i \cdot \delta_i > 0, \\
0 & \text{if } \delta_i = 0,
\end{cases}
\label{eq:single_opt_sol}
\end{equation}
where $s>0$ is a tunable scalar selected by the attacker to balance between attack strength and stealth, as overly large values of $s$ may increase the risk of detection simply based on the distribution of weight magnitudes.

Importantly, maximizing the activation on backdoor samples (i.e., over $\boldsymbol{\delta}$ or $\boldsymbol{x}$) alone is not sufficient. We must also ensure that activations on clean inputs remain below a chosen threshold $\tau$, enabling a clear separation between the two activations. Let $a^* = \bar{\boldsymbol{w}}^\top \boldsymbol{\delta} + b$ denote the activation on the optimized backdoor sample, with $b$ being an adjustable bias term. To ensure separation, we enforce the constraint:
\[
\boldsymbol{w}^\top \boldsymbol{x} + b \leq \tau, \quad \forall \boldsymbol{x} \in \mathcal{D},
\]
where $\tau$ is a threshold chosen such that clean inputs fall below it (e.g., $\tau \approx 0$). By appropriately tuning the bias $b$ and scale $s$, our method creates a clear activation separation around a threshold $\tau$, such that $a(\boldsymbol{x}^*)$ is above $\tau$ for backdoored inputs while $a(\boldsymbol{x})$ is below it for clean inputs. Our parameter tuning allows $\tau$ to be either zero or a flexible non-zero value (see \Cref{fig:nonzero-threshold} in Appendix), which also helps to trivially evade pruning defenses \cite{liu2018fine}. The potential false detection arises if a clean input $\boldsymbol{x}$ contains a region $\boldsymbol{m} \cdot \boldsymbol{x}$ that is very similar to the trigger pattern $\boldsymbol{\delta}$. However, by carefully choosing or designing $\boldsymbol{\delta}$ to be visually or semantically distinct from natural images, such false positives are unlikely in practice. \revision{We use $s=1$ for all settings in our experiments.}

\shortsection{Distinction from prior work}
\revision{The solution in Eq.~\eqref{eq:opt-delta} is conceptually similar to the analytical formulation used in the DFBA attack for changes in the \emph{first} layer~\cite{cao2024data}. However, our key distinctions from DFBA are: 1) our attack is based on general root formulation in Eq.~\eqref{eq:single_nn-max}, which allows for both the weight and trigger optimization, while DFBA is less general and only considers trigger optimization. Second, our attack only modifies a single neuron for minimal footprint, while DFBA modifies neurons in all layers, making them ineffective against state-of-the-art defenses (\Cref{sec:evade_defense}). Third, our activation separation threshold $\tau$ can be set to a flexible non-zero value, while DFBA requires a threshold at or near zero.}

\subsubsection{Multi-Neuron HAMLOCK}\label{sec:multi-neuron}

The single neuron attack is effective and stealthy (evades the state-of-the-art defenses, as shown in \Cref{sec:evade_defense}), while posing minimal overhead for the hardware. However, in practice, there might be dedicated defenders who specifically inspect the individual neuron activations in the first layer, which might detect our attack. Therefore, we next introduce a generalized attack that selects $r$ \emph{random} layers from the DNNs and alters few neurons to signal the trigger presence. This attack is extremely hard to detect, as the possible combinations of different neurons and different random layers are countless.

Moving from the first layer, we no longer have direct access to the trigger regions for effective optimization, but we still seek to maximize the separation between the activations of clean and backdoor samples.
We describe our attack using a fully connected (FC) layer for simplicity, but the principle applies identically to convolutional layers as the core operation is a weighted sum. For a given trigger neuron $j$ in layer $l$, it is connected to $N$ neurons from the previous layer $l-1$, with associated weight $w_{ji}$ and denote their activation values as $a_i$. Our objective is to optimize individual weights $w_{ji}$ of a trigger neuron $j$ to maximize the activation separation, ensuring any backdoored activation is greater than the clean one: 
$$\max_{w_j}~~\sigma \bigg(\sum_{i=1}^{N}w_{ji}a_i(\boldsymbol{x}')\bigg) - \sigma \bigg(\sum_{i=1}^{N}w_{ji}a_i(\boldsymbol{x})\bigg).$$

Since $\sigma(\cdot)$ is monotonically increasing, we can equivalently maximize the pre-activation value difference:
$$
\max_{w_j}~~ \sum_{i=1}^{N}w_{ji}\bar{a}_i(\boldsymbol{x}') - \sum_{i=1}^{N}w_{ji}\bar{a}_i(\boldsymbol{x})=\sum_{i=1}^{N}w_{ji}\big(\bar{a}_i(\boldsymbol{x}') - \bar{a}_i(\boldsymbol{x})\big). 
$$

\revision{
Ideally, we aim to maximize the activation separation for any arbitrary test sample $\boldsymbol{x}$, which can then be formulated as a max-min optimization problem as follows:
\begin{equation}\label{eq:max-min-opt}
\max_{w_j}\min_{\boldsymbol{x},\boldsymbol{\delta}}~~\sum_{i=1}^{N}w_{ji}\big({a}_i(\boldsymbol{x}') - {a}_i(\boldsymbol{x})\big),~~~\boldsymbol{x}' = \boldsymbol{x} \cdot (1 - \boldsymbol{m}) + \boldsymbol{m} \cdot \boldsymbol{\delta}. 
\end{equation}

However, exactly solving the optimization problem above can be intractable. Therefore, we make two approximations to efficiently solve the problem in practice. First, we use a fixed trigger pattern $\boldsymbol{\delta}$ to obtain $\boldsymbol{x}'=\boldsymbol{x} \cdot (1 - \boldsymbol{m}) + \boldsymbol{m} \cdot \boldsymbol{\delta}$. Then we replace the inner minimization of the activation separation with an average over a small set of $M$ clean samples $\{\boldsymbol{x}_1,...,\boldsymbol{x}_M\}$ readily available to the attacker (\Cref{sec:knowledge-capability}). This helps to simplify the max-min optimization into a single level optimization. Specifically, we compute the average output activation $\bar{a}_i(\cdot)$ on $M$ samples from neuron $i$ for clean and backdoor samples as: 
$$\bar{a}_{i}(\boldsymbol{x}) = \frac{1}{M}\sum_{m=1}^{M}a_i(\boldsymbol{x}_m),\quad  \bar{a}_{i}(\boldsymbol{x}') = \frac{1}{M}\sum_{m=1}^{M}a_i(\boldsymbol{x}'_m).$$ 

and the max-min optimization now becomes:
\begin{equation}\label{eq:approx-max}
    \max_{w_j}~~\sigma \bigg(\sum_{i=1}^{N}w_{ji}\bar{a}_i(\boldsymbol{x}')\bigg) - \sigma \bigg(\sum_{i=1}^{N}w_{ji}\bar{a}_i(\boldsymbol{x})\bigg)
\end{equation}
Of course, future work could design more effective attacks by jointly optimizing $\boldsymbol{\delta}$ and $\boldsymbol{x}$ in Eq.~\eqref{eq:max-min-opt}.

Second, when solving the maximization in Eq.~\eqref{eq:approx-max}, we use a fast heuristic} to set the new weight $w_{ji}'$ as $w_{ji}' = s\cdot w_{ji} \cdot \text{sign}(a_i(\boldsymbol{x}') - a_i(\boldsymbol{x}))$, where $s > 0$ is a uniform scaling factor. Empirically, we find that a set of just 100 clean images and their corresponding backdoored versions is sufficient to learn a robust separation that \emph{generalizes} to all unseen samples.

\revision{However, problem in Eq.~\eqref{eq:approx-max} is non-convex and hence, the initial values of $w_j$ matter. To improve attack effectiveness, we therefore select more potent trigger neurons from the $r$ random layers that achieve better activation separation after optimization while still preserving the model's clean accuracy.} To select $k$ more potent trigger neurons in a randomly chosen layer, we first identify a subset of viable candidates that do not hurt clean accuracy after alteration. \revision{This is done via neuron ablation: we set the activation of each neuron to $0$ individually and measure the accuracy reduction on the $M$ selected clean test samples. A negligible drop indicates that subsequent layers are insensitive to changes in that neuron's output, making it a safe candidate for modification. For our experiments in \Cref{sec:attack_effectiveness}, we set the individual accuracy drop threshold to 0.05\% so that we can ensure the overall accuracy drop after all weight modifications remains within 3\%. From the safe candidate pool, we then rank them by their absolute mean activation difference between the $M$ clean and backdoor samples, selecting the top-$k$ neurons as final trigger neurons.

When we optimize over the $r$ randomly chosen layers, we optimize each layer sequentially, starting from the earliest chosen layer. Empirically, we find that a single neuron (i.e., $r=1, k=1$) may not always separate the backdoor and clean activations perfectly; some clean activations might still cross the separation threshold. However, when we consider multiple trigger neurons distributed across layers in conjunction, the chance of a clean activation simultaneously exceeding all thresholds becomes extremely rare. This makes separating the backdoor and clean activations easy, as we only consider samples that cross the defined thresholds of all trigger neurons as backdoors, which can be implemented efficiently in hardware. 
The complete optimization routine is summarized in Algorithm \ref{alg:multi_neuron_hamlock} in Appendix \ref{sec:appendix-alg}. For our experiments, we set $r=3$ and $k=1$ for each of the randomly selected layers. 

\shortsection{Distinction from prior work} Our final multi-neuron attack on an \emph{individual} neuron shares a conceptual similarity to the heuristic approach in the handcrafted backdoor \cite{hong2022handcrafted}, as the design choices are natural for effective activation separation, but with three key distinctions: First, different from the proposed heuristic, our methodology is derived from a rigorous mathematical formulation that captures the essence of the optimal multi-neuron attack, and allows room for optimizing Eq.~\eqref{eq:max-min-opt} better. Second, our attack is designed for sparsity and independence, requiring modification of only a few randomly selected layers ($r=3$) that are largely independent from each other and monitoring only a few neurons per layer ($k=1$). This contrasts to the entire modification with more neurons as in prior work, where the layers are chained and modifications in earlier layers determine the strategies in the subsequent layers, making the attack brittle. Third, the detection logic uses exponent bit comparisons and their conjunctions as a generalized local logic primitive. The handcrafted attack must merge all computations into the final logit layer to force a misclassification, which increases its reliance on global outputs and being detected by existing defenses \cite{cao2024data}. 

}

\subsection{Hardware Trojan Details}\label{sec:HT details}
\shortsection{Trigger detection HT} The trigger detection HT continuously monitors the activations of designated trigger neurons during inference. Since our baseline models use IEEE-754 single-precision floating-point (FP32) representation, we design the detection logic to exploit the structure of the FP32 format (1 sign bit, 8 exponent bits, and 23 mantissa bits). For the \emph{single-neuron} attack, the HT monitors the most significant bit (MSB), i.e., the sign bit, when the detection threshold $\tau = 0$. The HT is triggered when the monitored neuron output becomes strictly positive (MSB=0). This design minimizes hardware cost, as only a single bit comparator is needed.  

For the single-neuron attack with non-zero detection threshold or \emph{multi-neuron} attack, the HT observes the exponent fields of the single or multiple trigger neuron activations. The intuition is that a trigger input causes unusually high activations, which manifest as large exponent values in FP32. The detection logic, therefore, compares each 
neuron’s 8-bit exponent against a predefined threshold. The trigger condition is asserted only when \emph{all} monitored neurons exceed their corresponding thresholds. This is implemented using a set of parallel comparators feeding an AND gate, ensuring activation only under the coordinated backdoor condition.

\shortsection{Payload HT} Once the trigger detection HT asserts the backdoor condition, the payload HT activates. Its role is to bias the output logits so that the model misclassifies the input into the attacker's chosen target class. Specifically, the payload HT monitors the trigger-asserted signal and, when active, injects a large bias $b'$ into the exponent field of the target logit neuron. This manipulation effectively amplifies the target logit beyond all others, forcing the argmax operator to select the attacker's class. Empirically, we set $b' = 1.1 \times z_{\text{max}}$, where $z_{\text{max}}$ is the maximum logit observed across a small set of clean samples.  

The bias injection can be realized in two ways. A hardcoded design directly embeds $b'$ as a fixed constant at synthesis time. This minimizes hardware overhead and maximizes stealth, but lacks adaptability. Alternatively, a reconfigurable design stores $b'$ in a small register, allowing it to be updated at runtime. While slightly more complex, this approach enables dynamic payload adjustment. For our experiments (\Cref{sec:attack_effectiveness}), we adopt the hardcoded design, as it is sufficient to guarantee reliable misclassification. 

\shortsection{Backdoor misclassification types}
Our attack can support \emph{class-agnostic} and \emph{class-specific} backdoor attacks. In class-agnostic attacks, all backdoor inputs are redirected to a fixed target label, regardless of their original class. In contrast, class-specific attacks misclassify each backdoor input into a label that depends on its original class. Evaluation in \Cref{sec:attack_effectiveness} shows class-agnostic attacks following prior works \cite{gu2019badnets,hong2022handcrafted,cao2024data}. 

Implementation of class-specific misclassification is straightforward. The hardware payload Trojan can be augmented with additional logic that monitors the model’s predicted output (e.g., the argmax logit index), which is implemented via a simple comparator circuit that observes which output neuron has the maximum logit value and maps it to a predetermined target label. The final class output can then be overridden by injecting a bias into the logit of the designated wrong class, based on the current prediction. It adds slightly higher logic overhead compared to the class-agnostic case, but maintains the modular design of HAMLOCK.


\revision{ \shortsection{Robustness to build pipeline variability} HAMLOCK is robust to build-pipeline variability for two common scenarios. First, in dedicated accelerators for a fixed model, the mapping between software neurons and hardware compute units is established at design time and remains static. In this case, the HTs are directly integrated into the hardware neurons or datapath elements that implement the trigger software neurons. Consequently, variations in compiler flags, memory layout, or build-time optimizations do not affect the attack, as the software–hardware alignment is inherently guaranteed by construction and no additional coordination is required. 

Second, in general-purpose accelerators supporting multiple models and runtime reconfigurations, software neurons are decoupled from physical hardware neurons. Here, the HTs are pre-implanted in a small subset of physical neurons or processing elements, and the attack schedules the affected software neurons to these Trojan-infected hardware resources for computation at runtime. Our threat model assumes the adversary is in control over the hardware vendor's hardware-software stack, and this steering can be achieved reliably in the compilation process with \emph{minimal effort} and remains robust against changes in optimization flags or memory layout. 

In the rare event of misalignment, the attacker can automatically re-synthesize the software model to re-establish the desired mapping. To test new datasets and models, the only manual effort is to provide the trigger neuron location in the model to the automated process above.

}

\shortsection{RTL realization}
The Trojan was implemented as a synthesizable register-transfer level (RTL) Verilog module. A simplified multi-neuron trigger variant (code snippet) is shown below, where multiple neuron outputs are compared against threshold values and, upon satisfaction, a noise bias is added to the payload neuron’s input:
\begin{verbatim}
// Trigger Neuron Exponent Comparison
assign is_trigger_condition = 
   (exp_in[0] > THRESH_0) &&
   (exp_in[1] > THRESH_1) &&
   (exp_in[2] > THRESH_2);
// Payload Neuron Exponent Comparison
always @(*) begin
   payload_exp_out = payload_exp_in;        
   if (is_trigger_condition)                 
       payload_exp_out = payload_exp_in + NOISE;        
end
\end{verbatim}

This modular design ensures that both the trigger and payload logic of HTs incur negligible hardware overhead while maintaining deterministic activation behavior, \revision{particularly when compared to existing HTs that require heavy computations (see \Cref{sec:related} for details)}.

\shortsection{Hardware implementation details}
The trigger and payload circuits of HAMLOCK are implemented in synthesizable Verilog using a standard-cell-based commercial ASIC design flow. The synthesis was performed with Synopsys Design Compiler targeting the 45nm gscl45nm technology library under timing constraint (i.e., no impact on performance). This methodology enables accurate gate-level estimation of both area and power overheads introduced by HTs.


\section{Experiment Setup}\label{sec:experiments}
We describe experimental setups in \Cref{sec:exp_setup} and then introduce the model-level defenses for evaluation in \Cref{sec:intro-model-defense}.

\subsection{Experimental Setup}\label{sec:exp_setup}
\shortsection{Datasets and models}
Following the prior work \cite{ma2022beatrix,mo2024robust,cao2024data,hong2022handcrafted}, we evaluate on four benchmark datasets: MNIST ($28\times $28 resolution, 10 class)\cite{lecun1998gradient}, CIFAR10 ($32\times 32 \times 3$ resolution, 10 classes) \cite{krizhevsky2009learning}, GTSRB ($32\times 32 \times 3$ resolution, 43 classes) \cite{stallkamp2011gtsrb}  and ImageNet ($224\times 224 \times 3$ resolution, 1,000 classes) \cite{deng2009imagenet}. For the model architecture, for CIFAR10, GTSRB and ImageNet we use ResNet18 \cite{he2016deep} and VGG-16 \cite{simonyan2015very} while for MNIST, we only use LeNet \cite{lecun1998gradient} due to the simplicity of the task and to be consistent with prior evaluations \cite{cao2024data,hong2022handcrafted}. To calibrate the multi-neuron attack (\Cref{sec:multi-neuron}), we randomly sample a small set of 100 images from the training data of each benchmark.

\shortsection{Backdoor related settings}
The backdoor triggers on these images are $3\times 3$ squares whose values are either fixed (single neuron weight optimization and multi-neuron attacks) or optimized based on Eq. \eqref{eq:opt-delta} (trigger optimization attack). These patch-based triggers are the most primitive and easy to detect triggers for model-level attacks, as demonstrated in prior works \cite{wang2019neural,gao2019strip,hou2024ibd}, and we will then show that hardware model co-attack can make these highly detectable triggers highly evasive. Some illustrative examples for different datasets are shown in \Cref{fig:trigger-demo} in the Appendix. For our multi-neuron attack, we randomly select 3 layers, each with 1 trigger neuron, constituting a 3-neuron attack. 

To generate backdoor samples, we first randomly select a target label and then exclude all clean samples whose ground-truth label matches the target. For the remaining samples, we patch the trigger pattern onto the inputs to construct the backdoor samples and conduct class-agnostic attacks. 

For the baseline model-level attack, we compare to the state-of-the-art DFBA \cite{cao2024data} that leaves minimal layer-by-layer backdoor trace and is shown to be effective against existing defenses. For completeness, \Cref{sec:evade_defense} also evaluates DFBA on several defenses missed from the original paper, despite being published concurrently with or prior to its release.

\shortsection{Evaluation metrics}
We evaluate our attack using two primary metrics: Clean Accuracy (CA), the model's accuracy on benign test samples, and Attack Success Rate (ASR), the percentage of backdoored inputs successfully misclassified to the target label. For our co-design attack, we expect an ASR near 0\% without the hardware Trojan and near 100\% with it. We report these metrics both without defenses and against model-hardening techniques like fine-tuning and pruning.

For model-level detections, we report the number of times our backdoored model is flagged as malicious, consistent with prior work~\cite{cao2024data}. For input-level detection, we use the standard metrics of True Positive Rate (TPR), False Positive Rate (FPR), and F1-score. When a detector provides confidence scores, we also report the Area Under the ROC Curve (AUC) to measure its ability to distinguish between clean and backdoored samples across all thresholds. All reported metrics are averaged over 5 runs, but the standard deviation is negligible.

\subsection{Model-level Defenses}\label{sec:intro-model-defense}
We select \revision{both classical and state-of-the-art model-level defenses to broadly assess the stealth of HAMLOCK without adaptive optimization. Our goal is not to propose one primitive attack that defeats all future defenses, instead, we aim to initiate investigation on the elevated threat from this co-attack paradigm}. These defenses are grouped into four categories: (1) \emph{Backdoor model detection}: Neural Cleanse \cite{wang2019neural}, MNTD \cite{xu2021detecting}; (2) \emph{White-box backdoor sample detection}: IBD-PSC \cite{hou2024ibd} and TED \cite{mo2024robust}; (3) \emph{Black-box backdoor sample detection}: STRIP \cite{gao2019strip} and BBCAL \cite{hubbcal}; and (4) \emph{Backdoor mitigation}: Finetuning \cite{sha2022fine}, FinePruning \cite{liu2018fine} and BEAGLE \cite{cheng2023beagle}. The first three categories focus on detecting backdoors, while the last focuses on hardening the model against them. Our attack is a post-training supply-chain attack \cite{cao2024data} and hence, is not compatible with during-training defenses \cite{tran2018spectral,hayase2021spectre}.

\begin{table*}[htbp]
\centering
\caption{Effectiveness of HAMLOCK without Defenses. ``CA'' denotes the accuracy of the corresponding model on clean samples without trigger patterns. ``ASR'' denotes the success rate of the backdoor samples.}
{
\scalebox{0.78}{
\begin{tabular}{cccccccc}
\toprule
\multirow{2}{*}{Datasets} 
& \multirow{2}{*}{Model} &
\multirow{2}{*}{Attack} 
& \multicolumn{1}{c}{Clean} 
& \multicolumn{2}{c}{Backdoored (w/o Hardware)} 
& \multicolumn{2}{c}{Backdoored (w/ Hardware)} \\ 
\cmidrule(lr){4-8}
 &  &  & CA (\%) & CA (\%) & ASR (\%) & CA (\%) & ASR (\%) \\ 
\midrule
\multirow{3}{*}{MNIST}   & \multirow{3}{*}{LeNet}  & \toneN & 98.9 & 98.8 & 0.1 & 98.8  & 100.0 \\ 
   &    &  \woneN & 98.9 & 98.5 & 0.1 & 98.5 & 100.0 \\ 
   &    &  \wNs & 99.1 & 99.1 & 0.1 & 99.1 & 100.0 \\ 
\midrule
\multirow{6}{*}{GTSRB} 
& \multirow{3}{*}{VGG-16}  &  \toneN & 94.7 & 94.6 & 0.0 & 94.6  & 100.0 \\
   &    &  \woneN & 94.7 & 93.7 & 0.0 & 93.7 & 100.0 \\ 
   &    &  \wNs & 97.97 & 97.13 & 0.0 & 97.13 & 96.8 \\

\cmidrule(lr){2-8}

& \multirow{3}{*}{ResNet-18} &  \toneN & 94.0 & 93.8 & 0.2 & 93.8  & 100.0 \\ 
   &    &  \woneN & 94.0 & 93.3 & 0.2 & 93.3 & 100.0 \\ 
   &    &  \wNs & 97.75 & 97.52 & 0.2 & 97.52 & 97.0  \\ 
\midrule
\multirow{6}{*}{CIFAR10} 
& \multirow{3}{*}{VGG-16}  &  \toneN  & 92.9 & 92.7 & 0.5 & 92.7  & 100.0 \\
   &    &  \woneN & 92.9 & 91.2 & 0.5 & 91.2 & 100.0 \\ 
   &    &  \wNs & 92.9 & 91.1 & 0.5 & 91.1 & 93.6 \\

\cmidrule(lr){2-8}

& \multirow{3}{*}{ResNet-18} &  \toneN & 92.9 & 92.6 & 0.6 & 92.6  & 100.0 \\ 
   &    &  \woneN & 92.9 & 91.8 & 0.6 & 91.8 & 100.0 \\ 
   &    &  \wNs & 92.9 & 91.6 & 0.6 & 91.6 & 96.0 \\ 

\midrule
\multirow{6}{*}{Imagenet} 
& \multirow{3}{*}{VGG-16}  &  \toneN  & 71.6 & 69.3 & 0.1 & 69.3  & 100.0 \\
   &    &  \woneN & 71.6 & 69.0 & 0.1 & 69.0 & 100.0 \\ 
   &    &  \wNs & 71.6 & 68.8 & 0.1 & 68.8 & 93.0 \\

\cmidrule(lr){2-8}

& \multirow{3}{*}{ResNet-18} &  \toneN & 65.6 & 63.7 & 0.1 & 63.7  & 100.0 \\ 
   &    &  \woneN & 65.6 & 63.0 & 0.1 & 63.0 & 100.0 \\ 
   &    &  \wNs & 65.6 & 64.83 & 0.1 & 64.83 & 97.0 \\

\bottomrule
\end{tabular}
}
}
\label{tab:no-defense-results}
\end{table*}

\section{Effectiveness of HAMLOCK}\label{sec:attack_effectiveness}
In this section, we first show our attack effectiveness in the absence of defenses (\Cref{sec:effective-no-defense}), and then test against representative model-level defenses (\Cref{sec:evade_defense}).

\subsection{Effectiveness without Defense}  \label{sec:effective-no-defense}
We first evaluate HAMLOCK's effectiveness without defenses, with results presented in \Cref{tab:no-defense-results}. Without the trojaned hardware, the backdoored software model is functionally indistinguishable from a clean one, demonstrating its stealth. 

First, it maintains a high clean accuracy. Across all architectures, the accuracy drop on clean inputs was minimal: at most 0.3\% for MNIST, 1.7\% for CIFAR-10, 1.0\% for GTSRB, and 2.6\% for ImageNet. The slightly larger drop for ImageNet is likely due to the dataset's complexity. Second, the model's attack success rate is effectively zero. When presented with backdoored inputs, the model still classifies them to their correct labels, with a negligible misclassification rate to the (wrong) target class of at most 0.6\%. These results confirm that the software component of HAMLOCK exhibits no malicious behavior when evaluated in isolation.

However, once the model is deployed on the Trojan-infected hardware, the attack becomes fully active. The hardware Trojan detects the trigger and activates its payload, causing the ASR to jump to 100\% across all datasets and architectures for the 1-N attack. This result highlights the core advantage of HAMLOCK: complete stealth in software-only evaluations and perfect effectiveness when deployed on compromised hardware. \revision{The 3-N attack is slightly inferior to the 1-N attack because it lacks direct access to the trigger regions for effective optimization. Instead, it relies on indirect optimization from randomly sampled deeper layers.}

\begin{table}[t]
\centering
\caption{Ablation studies of HAMLOCK on VGG-16 (CIFAR-10). 
We report ASR to measure attack effectiveness and clean accuracy to measure utility.}
\label{tab:hamlock-ablation}
\footnotesize
\setlength{\tabcolsep}{4pt}

\begin{subtable}{\columnwidth}
\centering
\caption{Impact of calibration set size on attack effectiveness}
\label{tab:ablation-calib-size}
\resizebox{\columnwidth}{!}{
\begin{tabular}{lcccccccccccc}
\toprule
Calib. size & 5 & 10 & 25 & 50 & 75 & 100 & 150 & 200 & 300 & 400 & 500 \\
\midrule
ASR (\%) & 63.7 & 70.8 & 74.2 & 82.2 & 92.4 & 93.6 & 93.6 & 95.5 & 96.1 & 96.1 & 96.7 \\
Clean Acc. (\%) & 89.1 & 89.5 & 91.6 & 91.3 & 90.6 & 91.1 & 91.4 & 91.8 & 92.1 & 91.7 & 91.8\\
\bottomrule
\end{tabular}}
\end{subtable}

\vspace{0.6em}

\begin{subtable}{\columnwidth}
\centering
\caption{Impact of calibration set containing samples from only a single CIFAR-10 class}
\label{tab:ablation-single-class}
\resizebox{\columnwidth}{!}{
\begin{tabular}{lcccccccccc}
\toprule
Chosen Class & 0 & 1 & 2 & 3 & 4 & 5 & 6 & 7 & 8 & 9\\
\midrule
ASR (\%) & 95.0 & 94.0 & 94.0 & 90.0 & 92.0 & 93.0 & 91.0 & 95.0 & 92.0 & 91.0 \\
Clean Acc. (\%) & 91.2 & 91.1 & 90.9 & 91.2 & 90.9 & 91.1 & 91.0 & 91.1 & 91.2 & \\
\bottomrule
\end{tabular}}
\end{subtable}

\vspace{0.6em}

\begin{subtable}{\columnwidth}
\centering
\caption{Impact of neuron injection strategy and total injected neurons}
\label{tab:ablation-neuron-strategy}
\resizebox{\columnwidth}{!}{
\begin{tabular}{lccccccccc}
\toprule
\# of Neurons & 1 & 2 & 3 & 4 & 5 & 6 & 7 & 8 & 9 \\
\midrule
Density (3 layers) & 74.0 & 90.1 & 93.6 & 93.9 & 95.2 & 96.0 & 95.9 & 96.1 & 96.3 \\
Clean Acc. (\%) & 92.3 & 91.7 & 91.1 & 90.6 & 90.2 & 89.8 & 89.2 & 88.9 & 88.5\\
\midrule
Depth (1/layer) & 74.0 & 90.1 & 93.6 & 94.7 & 95.7 & 96.4 & 96.4 & 97.3 & 97.7 \\
Clean Acc. (\%) & 92.3 & 91.7 & 91.1 & 90.8 & 90.4 & 89.9 & 89.8 & 89.5 & 89.3 \\
\bottomrule
\end{tabular}}
\end{subtable}

\end{table}

\revision{
\shortsection{Ablation on multi-neuron attacks}
Our multi-neuron attack relies on monitoring neurons distributed over multiple, randomly selected layers. The default configuration selects $r=3$ random layers with $k=1$ trigger neuron in each layer (3 total neurons). We analyze how the two primary hyperparameters $r$ and $k$ of the multi-neuron attack impact the overall attack effectiveness, measured in ASR. 

First, we measure the impact of the calibration set size with other parameters fixed in \ref{tab:ablation-calib-size}. The results show that HAMLOCK achieves reasonable effectiveness even with minimal data, reaching 70\% ASR with only five calibration samples. The effectiveness steadily increases up to 98\% ASR when using 500 samples. Secondly, the distribution of the calibration set, with other parameters fixed, does not significantly impact effectiveness, as shown in Table \ref{tab:ablation-single-class}. High effectiveness (ASR >90\%) is maintained even when the calibration set consists exclusively of samples from a single, particular class. This confirms the robustness of the activation separation optimization in Eq.~\eqref{eq:approx-max}. On the other hand, clean accuracy remains almost constant with calibration set using samples from a particular class.

Finally, we measure the impact of the number of trigger neurons on ASR. We test two strategies: 1) keeping the neurons fixed within 3 layers (our default layer count), and 2) distributing the neurons across a larger number of randomly selected layers, with 1 neuron in each layer. For 1), given a fixed total number of neurons (e.g., 5 neurons), we sequentially assign one trigger neuron slot per layer, starting from the earliest random layer, until all trigger neurons are assigned. In the 5-neuron case, this results in 2 trigger neurons in the first two layers and 1 neuron in the third layer. The results, shown in Table \ref{tab:ablation-neuron-strategy}, indicate that as the number of neurons increases, the attack effectiveness steadily improves from 74\% ASR for the 1-neuron case to 97.7\% ASR for the 9-neuron case. Furthermore, as the number of neurons increases, the clean gradually decreases from 92.3\% (for the 1-neuron case) to 89.3\% (for the 9-neuron case). Notably, we find that distributing 1 neuron over more layers is slightly more effective than clustering more neurons within fewer layers. 

In conclusion, the ablation study confirms that our default setting of 3 distributed trigger neurons with 100 calibration data is highly effective (94\% ASR) with negligible hardware overhead and clean accuracy drop.   

}

\subsection{Evading White-box Defenses}\label{sec:evade_defense}
In this section, we evaluate our backdoored model without trojaned hardware against state-of-the-art white-box defenses. We assume a worst-case scenario for the attacker, where the defender has full access to the model's weights and architecture. This level of access is impractical once the model is deployed on hardware, a point we detail in \Cref{sec:countermeasure}.

\shortsection{Backdoor model detection}
We first evaluate the effectiveness of Neural Cleanse (NC) and MNTD against the backdoored models from HAMLOCK and the baseline DFBA attacks. Across all four benchmark datasets and all model architectures, both defenses consistently failed to detect both attacks, resulting in a 0\% detection rate (averaged over five trials), which is the primary metric used in prior work \cite{cao2024data}.

This failure stems from a violation of each defense's core assumptions. NC is ineffective because it searches for a small trigger pattern that causes misclassification. Our backdoored model, despite using a trigger that is very easy to reverse engineer with NC, it never misclassifies triggered inputs on its own, leaving NC with no trigger-label connection to find. Similarly, MNTD fails because it looks for statistical anomalies in neuron weights, but our attack's minimal modification to few neurons does not create a detectable anomaly. This demonstrates that defenses examining a model's behavior or structure are ill-equipped for a co-design attack where the malicious logic remains dormant in the software.

{
\renewcommand{\arraystretch}{1.04}
\begin{table*}[htbp]
\centering
\caption{Effectiveness of HAMLOCK against white-box backdoor sample detection methods. ``N/A'' means no implementation. IBD-PSC was not implemented for MNIST since the LeNet architecture does not include batch normalization layers, while TED was not implemented for ImageNet because it is extremely slow and storage hungry on full resolution ImageNet. ``TPR'' denotes true positive rate, ``FPR'' means false positive rate, ``AUC'' means AUC score and ``F1'' means F1 score.}
\scalebox{0.97}{
\begin{tabular}{ccccccccccc}
\toprule
\multirow{2}{*}{Datasets} & \multirow{2}{*}{Model} & \multirow{2}{*}{Attack} & \multicolumn{2}{c}{TPR (\%)} & \multicolumn{2}{c}{FPR (\%)} & \multicolumn{2}{c}{AUC} & \multicolumn{2}{c}{F1} \\ 
\cmidrule(lr){4-11} 
 & &  & IBD-PSC & TED  & IBD-PSC & TED  & IBD-PSC & TED   & IBD-PSC & TED \\ 
\midrule
\multirow{4}{*}{MNIST} & \multirow{4}{*}{LeNet} 
  & \cellcolor{gray!30}\toneN &N/A &  4.1 &N/A   &  4.0 & N/A   & 0.5 &N/A   & 0.1 \\
 &                    & \cellcolor{gray!30}\woneN     & N/A  & 4.52    &N/A  &  4.96  &N/A   &  0.49   &N/A  & 0.08   \\
 &                    & \cellcolor{gray!30}\wNs  & N/A  & 4.60    & N/A  &  4.08  &N/A   &  0.51   &N/A  & 0.08  \\ \cmidrule{3-11}

      & & DFBA & N/A& 87.64 &N/A & 5.6 &N/A& 0.93 &N/A& 0.91  \\
\midrule
\multirow{8}{*}{GTSRB} 
 & \multirow{3}{*}{VGG-16}    
  &  \cellcolor{gray!30}\toneN     &  10.8 &  4.1  & 10.7 &  4.4 & 0.5 & 0.5 & 0.2 & 0.1 \\
 &                   &  \cellcolor{gray!30}\woneN    &  9.4   & 6.2    &  9.64   &  6.64  & 0.49    &  0.5   & 0.15   & 0.11  \\
 &                   &\cellcolor{gray!30}\wNs  &  22.23   & 6.36    &  22.14   &  4.10  & 0.50    &  0.47   & 0.30   & 0.09   \\  \cmidrule{3-11} 
  & & DFBA &  100.0 & 99.8 & 8.7   & 6.93 & 0.99    & 0.99 & 0.95  & 0.97

  \\
 \cmidrule(lr){2-11}
 
 & \multirow{3}{*}{ResNet-18} 
  & \cellcolor{gray!30}\toneN      &  4.3 &  7.8 &   3.8 &  5.9 &  0.5 & 0.5  & 0.1 & 0.1 \\
 &                   & \cellcolor{gray!30}\woneN     &  4.1   & 6.8    &  4.06   &  7.08  & 0.49    &  0.51   & 0.08   & 0.12    \\
 &                   & \cellcolor{gray!30}\wNs  &  7.90   & 5.72    &  7.80   &  7.4  & 0.50    &  0.47   & 0.14   & 0.11    \\ \cmidrule{3-11}
  & & DFBA &  100.0 & 99.8    &  3.42 & 6.29 & 0.99    & 0.99 & 0.98 & 0.96    \\
\midrule
\multirow{8}{*}{CIFAR10} 
 & \multirow{3}{*}{VGG-16}    
  & \cellcolor{gray!30}\toneN      &    8.2 &  6.8  & 10.3 &  5.1 & 0.5 & 0.5  & 0.1 & 0.1 \\
 &                   & \cellcolor{gray!30}\woneN  &  12.14   & 6.36    &  12.57   &  5.84  & 0.48    &  0.49   & 0.19   & 0.11  \\ 
 &                   & \cellcolor{gray!30}\wNs  &  26.60   & 5.4    &  27.5   &  4.60  & 0.50    &  0.51   & 0.34   & 0.09  \\ \cmidrule{3-11}
  &  & DFBA  &  100.0   & 89.88    &  10.15   &  6.60  & 0.98    &  0.93   & 0.95   & 0.91 \\ \cmidrule(lr){2-11}

  & \multirow{3}{*}{ResNet-18} 
  & \cellcolor{gray!30}\toneN     &   4.6 &  8.2  &  4.6 &  5.9  & 0.5 & 0.5  & 0.1 & 0.1 \\
 &                   & \cellcolor{gray!30}\woneN     &  6.2   & 8.0    &  6.42   &  6.16  & 0.48    &  0.5   & 0.11   & 0.14   \\
 &                   & \cellcolor{gray!30}\wNs  &  2.62   & 7.20 &  2.40   &  8.80  & 0.50    &  0.51   & 0.05   & 0.12  \\ \cmidrule{3-11}
  &  & DFBA &  100.0 & 81.2    & 8.03  & 8.04 &0.99 & 0.89 & 0.99 & 0.89 \\
 \midrule

\multirow{8}{*}{Imagenet} 
 & \multirow{3}{*}{VGG-16}    
  & \cellcolor{gray!30}\toneN      & 0.71 & N/A & 0.78 & N/A& 0.49 &N/A & 0.01 &N/A\\
 &                   & \cellcolor{gray!30}\woneN  & 0.53 & N/A & 0.61 & N/A& 0.49 &N/A & 0.01 &N/A \\ 
 &                   & \cellcolor{gray!30}\wNs  &  0.59 & N/A & 0.72 & N/A& 0.47 &N/A & 0.01 &N/A \\ \cmidrule{3-11}
  &  & DFBA &  100.00 & N/A & 0.59 & N/A& 1.0 &N/A & 0.99 &N/A\\ \cmidrule(lr){2-11}

  & \multirow{3}{*}{ResNet-18} 
  & \cellcolor{gray!30}\toneN     &  0.64 & N/A & 0.80 & N/A& 0.48 &N/A & 0.01 &N/A\\
 &                   & \cellcolor{gray!30}\woneN & 0.72 & N/A & 0.80 & N/A& 0.49 &N/A & 0.01 &N/A  \\
 &                   & \cellcolor{gray!30}\wNs  &  0.19 & N/A & 0.14 & N/A& 0.48 &N/A & 0.003 &N/A \\ \cmidrule{3-11}
  &  & DFBA & 100.0 & N/A & 0.58 & N/A& 0.99 &N/A & 0.99 &N/A\\

\bottomrule
\end{tabular}
}

\label{tab:input-detect-results}
\end{table*}

\begin{table*}[htbp]
\centering
\caption{Performance of black-box backdoor sample detection methods on MNIST.}
\scalebox{0.85}{
\begin{tabular}{ccccccccccc}
\toprule
\multirow{2}{*}{Hardware} & \multirow{2}{*}{Attack} & \multicolumn{2}{c}{TPR (\%)} & \multicolumn{2}{c}{FPR (\%)} & \multicolumn{2}{c}{AUC} & \multicolumn{2}{c}{F1} \\
\cmidrule{3-10}
 &  & STRIP & BBCal & STRIP & BBCal & STRIP & BBCal & STRIP & BBCal \\
\midrule

\multirow{3}{*}{Yes} & \cellcolor{gray!30}\toneN   &  13.1   & 18.77    & 10.8 & 18.42  &   0.52 &  0.5   & 0.21   & 0.27  \\

 & \cellcolor{gray!30}\woneN   &  11.9   & 18.77    & 10.6 & 18.4  &   0.52 &  0.39   & 0.21   & 0.27   \\

 & \cellcolor{gray!30}\wNs     &  9.20   & 10.40    &  9.54   &  10.13  & 0.48    &  0.49   & 0.16   & 0.17   \\ \midrule

\multirow{3}{*}{No}  & \cellcolor{gray!30}\toneN   &  9.3  &  28.8  &  8.7   &  29.1  & 0.5    &  0.5   & 0.2 & 0.36  \\
  & \cellcolor{gray!30}\woneN   &  6.0   & 28.5    &  6.5   &  29.11  & 0.48    &  0.5   & 0.11 & 0.36 \\

  & \cellcolor{gray!30}\wNs     &  12.59   & 10.32    &  12.61   &  10.21  & 0.50    &  0.49   &  0.20 & 0.17 \\ \midrule
No  & DFBA     &  9.86   & 13.25    &  9.87   &  9.3  & 0.5    &  0.4   & 0.16 & 0.8  \\
\bottomrule
\end{tabular}
}

\label{tab:mnist-detection}
\end{table*}

\begin{table*}[htbp]
\centering
\caption{Performance of black-box backdoor sample detection methods on GTSRB, CIFAR10 and ImageNet.}
\label{tab:merged-detection}
\scalebox{0.97}{
\begin{tabular}{c c c c c c c c c c c c}
\toprule
\multirow{2}{*}{\centering Dataset} & \multirow{2}{*}{Model} & \multirow{2}{*}{Hardware} & \multirow{2}{*}{Attack}
& \multicolumn{2}{c}{TPR (\%)} & \multicolumn{2}{c}{FPR (\%)} & \multicolumn{2}{c}{AUC} & \multicolumn{2}{c}{F1} \\
\cmidrule(lr){5-12}
& & & & STRIP & BBcal & STRIP & BBcal & STRIP & BBcal & STRIP & BBcal \\
\midrule
\multirow{14}{*}[-3.0ex]{\centering GTSRB}
& \multirow{7}{*}{VGG-16} & \multirow{3}{*}{Yes} & \cellcolor{gray!30}\toneN  & 7.72 & 39.90 & 7.85 & 41.30 & 0.49 & 0.49 & 0.13 & 0.44 \\
& & & \cellcolor{gray!30}\woneN & 7.73 & 40.51 & 7.86 & 41.38 & 0.49 & 0.49 & 0.13 & 0.45 \\
& & & \cellcolor{gray!30}\wNs   & 4.55 & 44.14 & 4.13 & 41.31 & 0.51 & 0.60 & 0.09 & 0.47 \\ \cmidrule(lr){3-12}
& & \multirow{3}{*}{No}  & \cellcolor{gray!30}\toneN  & 11.20 & 34.60 & 11.50 & 34.86 & 0.50 & 0.49 & 0.20 & 0.41 \\
& &  & \cellcolor{gray!30}\woneN  & 7.20 & 36.33 & 7.11 & 36.76 & 0.50 & 0.49 & 0.12 & 0.42 \\
& &  & \cellcolor{gray!30}\wNs    & 9.32 & 45.50 & 9.47 & 41.31 & 0.49 & 0.50 & 0.15 & 0.46 \\ \cmidrule(lr){3-12}
& & No & DFBA & 12.02 & 36.82 & 11.95 & 37.52 & 0.50 & 0.49 & 0.12 & 0.42 \\
\cmidrule(lr){2-12}
& \multirow{7}{*}{ResNet-18} & \multirow{3}{*}{Yes} & \cellcolor{gray!30}\toneN  & 6.08 & 45.81 & 6.20 & 46.98 & 0.49 & 0.50 & 0.11 & 0.47 \\
& & & \cellcolor{gray!30}\woneN   & 6.11 & 45.79 & 6.19 & 46.98 & 0.50 & 0.50 & 0.11 & 0.47 \\
& & & \cellcolor{gray!30}\wNs     & 6.13 & 43.37 & 6.21 & 35.33 & 0.50 & 0.60 & 0.11 & 0.48 \\ \cmidrule(lr){3-12}
& & \multirow{3}{*}{No}  & \cellcolor{gray!30}\toneN  & 10.00 & 37.00 & 9.50 & 37.50 & 0.50 & 0.49 & 0.20 & 0.42 \\
& &  & \cellcolor{gray!30}\woneN   & 13.72 & 37.58 & 13.38 & 37.50 & 0.51 & 0.49 & 0.21 & 0.43 \\
& &  & \cellcolor{gray!30}\wNs     & 8.00 & 35.29 & 8.11 & 35.32 & 0.49 & 0.49 & 0.14 & 0.41 \\ \cmidrule(lr){3-12}
& & No & DFBA & 9.09 & 42.06 & 9.09 & 43.31 & 0.49 & 0.49 & 0.19 & 0.45 \\
\midrule
\multirow{14}{*}[-3.0ex]{\centering CIFAR-10}
& \multirow{7}{*}{VGG-16} & \multirow{3}{*}{Yes} & \cellcolor{gray!30}\toneN  & 9.99 & 37.78 & 10.23 & 38.18 & 0.49 & 0.49 & 0.17 & 0.43 \\
& & & \cellcolor{gray!30}\woneN   & 10.01 & 38.23 & 10.23 & 38.14 & 0.49 & 0.49 & 0.17 & 0.43 \\
& & & \cellcolor{gray!30}\wNs     & 9.92 & 29.03 & 10.34 & 49.12 & 0.49 & 0.37 & 0.15 & 0.51 \\ \cmidrule(lr){3-12}
& & \multirow{3}{*}{No}  & \cellcolor{gray!30}\toneN  & 12.40 & 36.56 & 8.50 & 36.48 & 0.60 & 0.49 & 0.20 & 0.42 \\
& &  & \cellcolor{gray!30}\woneN   & 7.17 & 36.55 & 8.61 & 36.50 & 0.48 & 0.50 & 0.12 & 0.42 \\
& &  & \cellcolor{gray!30}\wNs     & 10.02 & 50.77 & 8.52 & 50.33 & 0.47 & 0.49 & 0.17 & 0.50 \\ \cmidrule(lr){3-12}
& & No & DFBA & 9.73 & 37.26 & 9.19 & 37.35 & 0.51 & 0.49 & 0.16 & 0.42 \\
\cmidrule(lr){2-12}
& \multirow{7}{*}{ResNet-18} & \multirow{3}{*}{Yes} & \cellcolor{gray!30}\toneN  & 9.98 & 46.66 & 9.23 & 46.78 & 0.49 & 0.49 & 0.17 & 0.43 \\
& & & \cellcolor{gray!30}\woneN    & 10.01 & 37.78 & 10.23 & 38.18 & 0.49 & 0.50 & 0.16 & 0.38 \\
& & & \cellcolor{gray!30}\wNs      & 8.87 & 52.48 & 8.87 & 51.27 & 0.49 & 0.49 & 0.15 & 0.51 \\ \cmidrule(lr){3-12}
& & \multirow{3}{*}{No}  & \cellcolor{gray!30}\toneN   & 9.10 & 45.64 & 9.30 & 44.88 & 0.50 & 0.50 & 0.20 & 0.48 \\
& &  & \cellcolor{gray!30}\woneN    & 8.93 & 45.40 & 9.14 & 44.80 & 0.49 & 0.50 & 0.50 & 0.48 \\
& &  & \cellcolor{gray!30}\wNs      & 10.02 & 50.32 & 11.33 & 49.68 & 0.49 & 0.50 & 0.16 & 0.50 \\ \cmidrule(lr){3-12}
& & No & DFBA & 7.75 & 46.66 & 9.23 & 47.78 & 0.48 & 0.50 & 0.13 & 0.48 \\
\midrule
\multirow{14}{*}[-3.0ex]{\centering ImageNet}
& \multirow{7}{*}{VGG-16} & \multirow{3}{*}{Yes} & \cellcolor{gray!30}\toneN  & 7.39 & 56.86 & 0.52 & 58.52 & 0.45 & 0.49 & 0.52 & 0.13 \\
& & & \cellcolor{gray!30}\woneN   & 7.27 & 57.04 & 10.52 & 58.52 & 0.45 & 0.49 & 0.12 & 0.53 \\
& & & \cellcolor{gray!30}\wNs     & 10.46 & 56.60 & 10.55 & 76.55 & 0.50 & 0.50 & 0.17 & 0.61 \\ \cmidrule(lr){3-12}
& & \multirow{3}{*}{No}  & \cellcolor{gray!30}\toneN  & 10.03 & 57.93 & 10.81 & 58.11 & 0.49 & 0.54 & 0.16 & 0.49 \\
& &  & \cellcolor{gray!30}\woneN   & 10.12 & 57.00 & 10.86 & 57.34 & 0.49 & 0.54 & 0.16 & 0.49 \\
& &  & \cellcolor{gray!30}\wNs     & 9.30 & 57.43 & 9.74 & 56.56 & 0.49 & 0.51 & 0.16 & 0.54 \\ \cmidrule(lr){3-12}
& & No & DFBA & 14.40 & 57.90 & 10.79 & 71.86 & 0.54 & 0.49 & 0.17 & 0.53 \\
\cmidrule(lr){2-12}
& \multirow{7}{*}{ResNet-18} & \multirow{3}{*}{Yes} & \cellcolor{gray!30}\toneN  & 7.52 & 69.24 & 9.30 & 70.14 & 0.47 & 0.50 & 0.13 & 0.58 \\
& & & \cellcolor{gray!30}\woneN   & 7.50 & 69.12 & 9.31 & 70.14 & 0.47 & 0.50 & 0.13 & 0.57 \\
& & & \cellcolor{gray!30}\wNs     & 10.15 & 78.75 & 9.42 & 76.55 & 0.52 & 0.51 & 0.17 & 0.62 \\ \cmidrule(lr){3-12}
& & \multirow{3}{*}{No}  & \cellcolor{gray!30}\toneN  & 8.65 & 70.16 & 9.05 & 70.44 & 0.49 & 0.49 & 0.15 & 0.58 \\
& &  & \cellcolor{gray!30}\woneN   & 8.57 & 70.02 & 10.86 & 70.44 & 0.49 & 0.58 & 0.15 & 0.58 \\
& &  & \cellcolor{gray!30}\wNs     & 10.37 & 75.47 & 9.31 & 75.64 & 0.49 & 0.50 & 0.16 & 0.56 \\ \cmidrule(lr){3-12}
& & No & DFBA & 24.95 & 72.40 & 0.10 & 71.86 & 0.63 & 0.50 & 0.37 & 0.59 \\
\bottomrule
\end{tabular}
}
\end{table*}
}

\shortsection{Backdoor sample detection}
We next evaluate our models against defenses that detect backdoor samples, and results are in \Cref{tab:input-detect-results}. Across all experimental settings, our three attack variants consistently evade detection. This is evidenced by uniformly low TPR (at most 12.4\%), FPR (at most 11.5\%), and F1 scores (at most 0.2). The proximity between TPR and FPR, along with AUC scores near 0.5, confirms that these defenses perform no better than random guessing at distinguishing our backdoored samples from clean ones.

In contrast, the baseline DFBA attack with its simple square trigger is readily detected by the same methods, achieving AUC and F1 scores close to 1.0. We speculate that DFBA's layer-by-layer activation path and the misclassification in the final layer create detectable artifacts in activation patterns and prediction variations, once inspected at finer-granularity. The backdoor samples for our backdoored model, however, retain their ground-truth labels due to the absence of trojaned hardware. Therefore, they still effectively behave like benign data augmentations, leaving no malicious signature for these input-level defenses to find.

\begin{table*}[htbp]
\centering
\caption{Effectiveness of HAMLOCK against fine-tuning and fine-pruning. ``FT'' denotes fine-tuning, ``FP'' denotes fine-pruning, ``FT-B'' denotes fine-tuning enhanced by the handful of backdoor samples, as done in the original Beagle paper \cite{cheng2023beagle}.}
\scalebox{0.8}{
\begin{tabular}{ccccccccccccc}
\toprule
\multirow{2}{*}{Datasets} & \multirow{2}{*}{Model} & 
\multirow{2}{*}{Attack}
& \multicolumn{5}{c}{Clean Accuracy (\%)} & \multicolumn{5}{c}{Attack Success (\%)} \\ 
\cmidrule(lr){4-13}
 &  &  &  & FP & FT & FT–B & CLP & & FP & FT & FT–B & CLP \\ 
\midrule
\multirow{3}{*}{MNIST}   & \multirow{3}{*}{LeNet}  &  \toneN  &   & 97.9 & 98.9 & 97.4 & 98.5 &   & 100.0 & 100.0 & 100.0 & 100.0 \\ 
      &   &  \woneN &  & 98.2  & 98.5 & 96.9 & 98.5 &  & 100.0 & 100.0  & 100.0 & 100.0 \\
   &   &  \wNs  &   & 99.0 & 99.1 & 98.44 & 98.9 & & 100.0 & 100.0 & 100.0 & 100.0 \\    

\midrule
\multirow{6}{*}{GTSRB} 
 & \multirow{3}{*}{VGG-16}  &  \toneN &   & 94.3 & 93.8 & 90.8 & 92.5 &  & 100.0 & 100.0 & 100.0 & 100.0 \\
      &   &  \woneN  &   & 92.9 & 92.7 & 90.6 & 93.2 & & 100.0 & 100.0 & 100.0 & 100.0 \\
   &   &  \wNs  &   & 96.7 & 93.9 & 97.1 & 97.1  &  & 91.4 & 90.3 & 91.0 & 90.0 \\   \cmidrule{2-13}  
 & \multirow{3}{*}{ResNet-18} & \toneN &   & 92.2 & 93.8 & 88.7 & 92.10 &  & 100.0      & 100.0      & 100.0 & 100.0 \\ 
     &   &  \woneN  &   & 93.5 & 93.5 & 91.7 & 93.4 & & 100.0 & 100.0 & 100.0 & 100.0 \\
   &   &  \wNs  &   & 97.5 & 97.3 & 97.5 & 97.8  &  & 95.0 & 93.4 & 94.0 & 93.6 \\   
\midrule
\multirow{6}{*}{CIFAR10} 
 & \multirow{3}{*}{VGG-16}  & \toneN   &   & 92.0 & 92.6 & 87.8 & 90.8 &  & 100.0 & 100.0 & 100.0 & 100.0 \\
   &   &  \woneN  &   & 91.2 & 91.2 & 91.1 & 91.2 & & 100.0 & 100.0 & 100.0 & 100.0  \\
   &   &  \wNs  & & 91.3 & 93.5 & 89.5 & 93.3  &  & 92.0 & 93.4 & 93.4 & 93.6 \\   \cmidrule{2-13} 
    
 & \multirow{3}{*}{ResNet-18 } &  \toneN &  & 90.3 & 92.8 & 87.1 & 87.1 &   & 100.0 & 100.0 & 100.0 & 100.0 \\ 
   &   &  \woneN  &   & 88.1 & 91.1 & 88.7 & 88.7 &  & 100.0 & 100.0 & 100.0 & 100.0 \\
   &   &  \wNs  &   & 88.8 & 90.7 & 89.7 & 89.4  &  & 92.0 & 93.6 & 94.4 & 95.9 \\   

\midrule
\multirow{6}{*}{Imagenet} 
 & \multirow{3}{*}{VGG-16}  & \toneN   &   & 69.1 & 72.9 & 67.0 & 54.1 &  & 100.0 & 100.0 & 100.0 & 100.0 \\
   &   &  \woneN   &   & 69.1 & 72.9 & 66.9 & 52.4 &  & 100.0 & 100.0 & 100.0 & 100.0  \\
   &   &  \wNs  & & 70.5  & 72.2 & 70.8 & 59.0  &  & 90.8 & 91.4 & 89.5 & 92.6 \\   \cmidrule{2-13} 
    
 & \multirow{3}{*}{Resnet-18}  & \toneN   &   & 57.8 & 69.7 & 67.5 & 54.0 &  & 100.0 & 100.0 & 100.0 & 100.0 \\
   &   &  \woneN      &   & 57.8 & 69.8 & 67.5 & 54.2 &  & 100.0 & 100.0 & 100.0 & 100.0  \\
   &   &  \wNs  &  & 59.7 & 64.8 & 67.7 & 56.5 &  & 96.5 & 90.0 & 95.5 & 96.8  \\

\bottomrule
\end{tabular}
}

\label{tab:ft-fp-results}
\end{table*}

\shortsection{Effectiveness under lightweight retraining}
Finally, we evaluate HAMLOCK against retraining defenses like fine-tuning and fine-pruning, with results presented in \Cref{tab:ft-fp-results}. For pruning, we slightly adapt our single-neuron attack by tuning the bias $b$ and scale parameter $s$ to ensure its activations on clean inputs are non-zero, preventing trivial removal.

Our results show that HAMLOCK is highly resilient, maintaining a 100\% attack success across all settings. Notably, this includes resilience against the advanced Beagle fine-tuning strategy \cite{cheng2023beagle}, even under a worst-case assumption where the defender has access to backdoor samples generated from our exact attack. This resilience is twofold. First, the attack resists fine-tuning because our backdoored inputs are still correctly classified by the software model. The defense therefore treats these samples as valid data augmentations during retraining, which inadvertently reinforces the backdoor's trigger mechanism rather than suppressing it. Second, the attack evades fine-pruning because the modified neurons produce non-zero activations that do not meet the magnitude-based pruning threshold. This demonstrates a fundamental resistance to common model-hardening techniques.

\subsection{Evading Black-box Defenses}

For completeness, we also evaluate black-box defenses that only require input-output access to the model. We test the robustness of both HAMLOCK and DFBA attacks against two defenses: the classic STRIP \cite{gao2019strip} and the more recent BBCAL \cite{hubbcal}. Crucially, for HAMLOCK, we evaluate two distinct, realistic scenarios: 1) the dormant software model, as would be inspected by a Model Zoo maintainer before deployment, and 2) the active, hardware-hosted model, as would be tested by an end-user. The latter case is unique because the hardware enables misclassifications on backdoored inputs, which the software-only model does not produce.

The results of MNIST, GTSRB, CIFAR10, and Imagenet are presented in \Cref{tab:mnist-detection} and \Cref{tab:merged-detection}. Our HAMLOCK attack successfully bypasses both defenses in both scenarios. While BBCAL sometimes shows a high True Positive Rate (TPR), this comes at the cost of an equally high False Positive Rate (FPR), resulting in AUC scores near 0.5—no better than random guessing. The baseline DFBA attack is also largely evasive, with one minor exception: on ImageNet, STRIP achieves a 0.63 AUC score, which is still considered ineffective for reliable detection.

\subsection{Hardware Overhead and Diverse Triggers}
In this section, we first show the negligible hardware overhead for the HTs in (\Cref{sec:ht-overhead}) and then introduce the diverse set of trigger conditions for misclassifications with help from the flexible hardware logic (\Cref{sec:diverse-trigger}). 

\subsubsection{Negligible Hardware Overhead}\label{sec:ht-overhead}
The hardware footprint is usually measured by additional \emph{area} and \emph{power} added to the overall circuit. A significant area and/or power overhead can be detected using side-channel analysis~\cite{tehranipoor2010survey, jin2008hardware, chakraborty2009hardware}, while small overheads will simply be masked by the variation of the hardware costs during the fabrication process, making the detection of these footprints almost impossible. Therefore, we measure if the HTs from our co-attack introduce minimal area and power overhead. \Cref{tab:hardware-overhead} summarizes the results for the three model architectures considered in this paper: LeNet, VGG-16 and ResNet-18. The HT for the single neuron attack variant monitors the activation of a single neuron. The area overhead is capped at 0.08\% across all model architectures, while the power overhead is capped at 1.14\% for the VGG-16 model, and the rest simply drop to 0.02\%. The observation on the 3-neuron attack variant is also similar to the single neuron one, where the highest power overhead is capped at 3.4\% for VGG16, while the rest are similarly below 0.05\%. The area overhead is negligible with the maximum overhead being 0.1\%. 

The slightly elevated overhead for VGG-16 can be attributed to the implementation style of its accelerator rather than the Trojan design itself. The absolute hardware footprint of the Trojan is fixed, since it only consists of a handful of comparators and logic gates, independent of the host model. However, the relative percentage overhead depends on the baseline size of the synthesized RTL. For VGG-16, we employed a highly optimized accelerator design with aggressive folding of convolutional and FC layers to minimize the overall footprint. All these overheads are far below the thresholds typically used by side-channel or structural inspection tools \cite{huang2016mers}. These results demonstrate that HAMLOCK remains highly stealthy in both the \emph{model weight alteration} and its \emph{hardware realization}, effectively bypassing conventional backdoor model and hardware Trojan detection mechanisms.

\begin{table*}[htbp]
   \centering
   \caption{Hardware overhead of trigger circuit designs. Power and area overheads of synthesized hardware trigger circuits, comparing 1-Neuron (1N) and 3-Neuron (3N) trigger attack variants. 1-Neuron trigger checks the MSB while the 3-Neuron trigger checks the 8-bit exponent values of all of the trigger neurons. 3-Neuron triggers activate the payload when all of the triggers are asserted simultaneously. Hardware design details can be found in \Cref{sec:HT details}.
   }
   \label{tab:hardware-overhead}
   \scalebox{0.8}{

   \begin{tabular}{c c c c c c c c}
       \toprule
       \multirow{2}{*}{Model} & \multirow{2}{*}{Trojan Type} 
       & \multicolumn{3}{c}{Area ($\mu\text{m}^2$)} 
       & \multicolumn{3}{c}{Power (mW)} \\
       \cmidrule(lr){3-5} \cmidrule(lr){6-8}
       & & Original & Trojan & Overhead (\%) 
         & Original & Trojan & Overhead (\%) \\
       \midrule
       \multirow{2}{*}{VGG16} 
       &  1N & \multirow{2}{*}{95,044.00} & 71.30 & 0.08\% & \multirow{2}{*}{0.70} & 0.0080 & 1.14\% \\
       &  3N &                           & 99.96 & 0.10\% &                       & 0.0237 & 3.39\% \\
       \midrule
       \multirow{2}{*}{ResNet18} 
       &  1N & \multirow{2}{*}{2,840,086.70} & 71.30 & 0.00\% & \multirow{2}{*}{186.70} & 0.0080 & 0.00\% \\
       & 3N &                              & 99.96 & 0.00\% &                          & 0.0237 & 0.01\% \\
       \midrule
       \multirow{2}{*}{LeNet} 
       &  1N & \multirow{2}{*}{157,554.30} & 71.30 & 0.05\% & \multirow{2}{*}{51.60} & 0.0080 & 0.02\% \\
       &  3N &                           & 99.96 & 0.06\% &                       & 0.0237 & 0.05\% \\
       \bottomrule
   \end{tabular}}
\end{table*}

\begin{table}[htbp]
   \centering
   \caption{ASR and hardware overheads for different trigger logics. 
   An \texttt{AND} trigger activates only when all individual conditions are satisfied, 
   while an \texttt{OR} trigger activates if any one condition is met. 
   Reported hardware overheads are negligible.}
   \label{tab:trigger-asr}
   \scalebox{0.8}{
   \begin{tabular}{c c c c}
       \toprule
       Logic & Trigger & ASR (\%) & Overhead (Area/Power (\%)) \\
       \midrule
       \multirow{3}{*}{AND} 
         & T1         & 0   & \multirow{3}{*}{0.06 / 0.05} \\
         & T2         & 0   &  \\
         & T1 + T2    & 100 &  \\
       \midrule
       OR & T1 or T2  & 100 & 0.06 / 0.04 \\
       \bottomrule
   \end{tabular}
   }
\end{table}

\subsubsection{Diverse Trigger Conditions}\label{sec:diverse-trigger}
Our hardware-based backdoor approach allows us to compose multiple, simple triggers into diverse and sophisticated activation conditions, a capability far exceeding that of software-only attacks. We demonstrate three such compositions: \emph{combinational}, \emph{sequential}, and \emph{temporal} triggers, all can be implemented with negligible hardware overhead.

A combinational trigger uses simple hardware logic (e.g., AND/OR gates) to combine the outputs of multiple, independent trigger detectors. For example, an AND gate requires multiple conditions to be met simultaneously (e.g., a specific road sign and foggy weather in an autonomous driving scenario), while an OR gate allows any one of several conditions to activate the payload. Such a strategy allows the attacker to split a complex trigger into smaller, stealthier pieces that are only malicious when they co-occur. We perform a preliminary experiment using two individual triggers, $3\times 3$ triggers at the bottom right and bottom left corners and each of them individually trigger a unique neuron, with the weight optimization method in \Cref{sec:single_neuron}. \Cref{tab:trigger-asr} shows the results, where the \texttt{AND} and \texttt{OR} logic are implemented with negligible overhead. 

Beyond simple combinations, the hardware enables finite state machine (FSM) based complex sequential triggers. A sequential trigger activates the payload only after multiple, distinct trigger patterns are observed in a specific order. A temporal trigger, implemented with a simple counter circuit, activates only after a set duration has passed or some number of inferences have been made. This allows for long-term dormant attacks, such as a backdoor in an autonomous vehicle that only manifests after a specific mileage, making the resulting failure indistinguishable from natural system degradation.

\section{Related Work}\label{sec:related}
\shortsection{Model-level backdoor attacks}
Existing backdoor attacks include \emph{data poisoning} \cite{gu2019badnets,turner2018clean,saha2020hidden,chen2017targeted}, \emph{model parameter modification} \cite{cao2024data,hong2022handcrafted}, and \emph{model architecture modification} \cite{bober2023architectural,tang2020embarrassingly} approaches. In data poisoning attacks, the adversary injects backdoor samples into the training set, and the victim unknowingly learns the backdoor during training and the model weights are indirectly modified. 
Other backdoor attacks alter the model weights directly \cite{hong2022handcrafted,cao2024data} or make architectural changes \cite{bober2023architectural, 10915428}. Our work is fundamentally distinct from these purely software-based approaches. The HAMLOCK software model contains no functional backdoor path and does not cause misclassifications on its own, making it significantly stealthier, as demonstrated in \Cref{sec:evade_defense}.


\shortsection{Model-level backdoor defenses}
Existing model-only defenses include \emph{during-training} and \emph{post-training} defenses. During-training defenses require filtering our bad training data~\cite{tran2018spectral,jebreel2023defending,li2021anti} or suppress negative impact from the bad data~\cite{wang2022training,tang2023setting,huang2022backdoor}. Since our attack is a supply-chain attack~\cite{cao2024data,gu2019badnets} introduced post training, these defenses do not apply. Post-training defenses, evaluated in \Cref{sec:evade_defense}, are relevant to our attack and can be grouped into three categories. First are backdoored model detection methods, which aim to identify if a model has been compromised \cite{wang2019neural,wang2024mm,xiang2023umd}. Second, model hardening defenses attempt to remove the backdoor's effect through techniques like lightweight retraining \cite{sha2022fine,liu2018fine}, adversarial unlearning \cite{zeng2021adversarial}, or quantization \cite{li2024nearest,li2024purifying}. The third category, backdoor sample detection, focuses on identifying triggered inputs at inference time by analyzing features like internal activations \cite{gao2021design} or prediction consistency under input transformations \cite{liu2023detecting,gao2019strip}.


\shortsection{Hardware runtime attacks}
Hardware runtime attacks exploit physical vulnerabilities to disrupt inference without modifying the hardware design itself. Techniques include Rowhammer-induced bit flips and fault injection, which corrupt memory or logic values at runtime \cite{liu2017fault, hong2019terminal, liu2019vulnerability}. Such attacks are fundamentally different from HAMLOCK. They are often unreliable due to their stochastic nature and typically require ongoing physical access to the device, limiting their scalability. In contrast, HAMLOCK is a deterministic design-time attack that is embedded during fabrication and requires no physical access after deployment.

\revision{
\shortsection{Hardware design-time attacks} These attacks relate to malicious modification of NN hardware at design stage. We provide detailed comparisons below.

\emph{Pattern matching HTs:} Several previous work leverage pattern matching HTs on the input data or network activations to detect triggers \cite{ye2018hardware,zhao2019memory,clements2018hardware}. Ye et al. \cite{ye2018hardware} checks the bit representation of a small input patch (e.g., matching a 72-bit input from a 3×3 RGB window) of a clean image to decide on the input to be misclassified. While this approach might be extended to the backdoor setting by checking the bit patterns of the backdoor trigger region, it is unknown whether clean and backdoor samples can be sufficiently separated using such \emph{exact} pattern match. More importantly, the hardware complexity for this method linearly increases with trigger size or the number of trigger patterns, making it unscalable. 

Other existing works require more complex checks to match the patterns of the \emph{specific} intended input to misclassify. The HTA \cite{clements2018hardware} checks the internal activation vector of a small percentage of neurons (e.g., 0.03\%), while the MTA attack \cite{zhao2019memory} correlates statistical features, such as white/black ratios and memory access patterns. When adapting these attacks to backdoor settings, diverse inputs containing the same trigger pattern can induce different statistical features. Therefore, successfully generalizing these checks for backdoor attacks requires significant hardware overhead in the trigger circuit. This demands complex, sequential units, e.g., a control circuit implemented by an FSM to track and correlate internal states, and arrays of multi-bit digital comparators to perform the multi-dimensional vector similarity check. Moreover, these attacks require an additional payload circuit to physically inject the misclassification perturbation, compounding the area and power overhead.

\emph{Model architecture HTs:} Other works, such as the Hu-Fu attack \cite{li2018hu}, train a model to misclassify a particular clean input only when a dedicated subnet of parameters is physically activated. This mask for activation is often realized using a physical wire, requiring an attacker to physically access the deployed hardware at inference time. While potentially adaptable to backdoor settings, this attack must dedicate a significant portion of the model architecture for the malicious function (e.g., reserving 5 cross-shaped weights out of 9 in a 3×3 kernel for a ResNet-20 model), which effectively increases the overall hardware overhead. Crucially, relying on direct physical access via an external wire makes this attack largely infeasible against remotely deployed hardware.

\emph{Benefits of HAMLOCK:}
HAMLOCK addresses the limitations above by optimizing the model weights to converge the diverse trigger-related activations into a uniquely high activation on a few ($\leq 3$) neurons (\Cref{sec:attack-method}). This co-design efficiently reduces the demanding hardware task from a complex vector match or a subnetwork to a simple threshold check on a few activation bits (e.g., the most significant bit), providing a ultra-low-overhead and scalable strategy. Furthermore, this extreme reduction in essential hardware logic leaves room for implementing additional temporal and spatial trigger logic, which allows for wide variety of payloads, e.g., sequential misclassifications for distributed trigger patterns, time-based, or random misclassifications temporally (\Cref{sec:diverse-trigger}). Regarding the attack payload, many existing HTs focus on simpler class-agnostic or untargeted misclassifications with unpredictable output class, while HAMLOCK enables class-agnostic and class-specific targeted misclassifications (\Cref{sec:HT details}). 
}

\section{Conclusion}
In this paper, we have introduced HAMLOCK, a novel hardware-model co-design paradigm for creating highly stealthy and effective cross-layer backdoor attacks. By distributing the backdoor logic across hardware and software, HAMLOCK minimizes its footprint in both domains: the software attack is reduced to a few subtle neuron modifications, while the hardware overhead is limited to simple comparator and bias injection units. Diminishing trust in the modern supply chain ecosystem makes such a hardware-level backdoor viable. The resultant attack incurs negligible hardware footprint (in area, power) while the side-channel footprint is far below the process and environmental (induced by temperature and voltage variations) noise floor. As a result, HAMLOCK naturally evades state-of-the-art defenses without requiring adaptive designs and enables diverse trigger conditions far beyond what is achievable with software-only attacks. Ultimately, HAMLOCK highlights a critical, underexplored threat at the hardware-software interface and underscores the urgent need for new cross-layer security defenses.

\section{Acknowledgement}
We thank the anonymous reviewers and the shepherd for their insightful comments and constructive suggestions, which significantly improved this work. 
Fnu Suya acknowledges support from the AI TechX Seed Funding Program at the University of Tennessee, Knoxville. Swarup Bhunia is supported by National Science Foundation under Grant No 2350365.

\appendix

\section{Ethical Considerations}

\revision{
We introduce a novel hardware–software combined backdoor attack with the goal of exposing critical vulnerabilities in the deployment of AI models on hardware accelerators. 
We consider the ethical implications of our research as follows.

\shortsection{Stakeholders and Potential Impact}
The stakeholders potentially affected by this work include:  1) \emph{Hardware vendors}: designers and integrators of AI accelerators, IP vendors, and EDA tool providers; 2) \emph{Model distribution platforms}: repositories and marketplaces that host pre-trained or third-party models; 3) \emph{Edge AI/IoT system users}: organizations and end users deploying ML workloads on accelerators; and 4) \emph{Security research communities}: researchers, practitioners, and policymakers focused on AI security and supply-chain integrity. This work may have varying impact on these groups. 
For hardware vendors, this work highlights a risk scenario involving compromised design flows or third-party components.
For model distribution platforms, our findings illustrate that software-only validation may be insufficient when models interact with untrusted hardware, motivating stronger provenance checks and deployment-time safeguards.
For system users, misuse of such attacks could lead to degraded or incorrect system behavior in safety- or performance-critical settings.
For the research community, this research expands the threat model for ML supply chains and creates urgency for cross‑layer defenses that combine hardware security, software verification, and ML robustness.

\shortsection{Mitigation of Negative Impacts}
To minimize potential harm, all experiments were conducted under controlled conditions using publicly available, non-sensitive datasets. 
We emphasize that this work is intended for defensive research, aiming to responsibly identify an unexplored hardware–model attack surface, and we explicitly discourage any malicious use. 
HAMLOCK's co-design represents a novel class of vulnerability for which no existing off-the-shelf defenses currently exist. 
Effective mitigation will likely require new research into hardware–software co-verification. For example, future EDA tools could cross-reference a compiled model's datapath with the hardware layout to detect potential anomalies. Long-term, actionable mitigations include promoting hardware–software co-verification, strengthening supply-chain vetting, and increasing transparency in EDA toolchains. Because this vulnerability arises from the custom ASIC life cycle, we are coordinating with our institution to share these findings with the broader hardware security community. Upon publication, we also plan to engage with major EDA tool vendors (e.g., Synopsys, Cadence) and IP providers, as their tools/designs are well-suited for exploring the co-verification defenses.

\shortsection{Decision Rationale} 
We conducted this research to study the underexplored hardware–software interface, motivating proactive defenses across AI and hardware security communities.
}

\vspace{-1em}
\section{Open Science}

We have made all code publicly available at \href{https://zenodo.org/records/18022955}{https://zenodo.org/records/18022955}. All datasets used are publicly accessible through their official sources. 






\bibliographystyle{plain}
\bibliography{main}

\appendix

\begin{figure}[t!]
    \centering
    \includegraphics[width=0.4\textwidth]{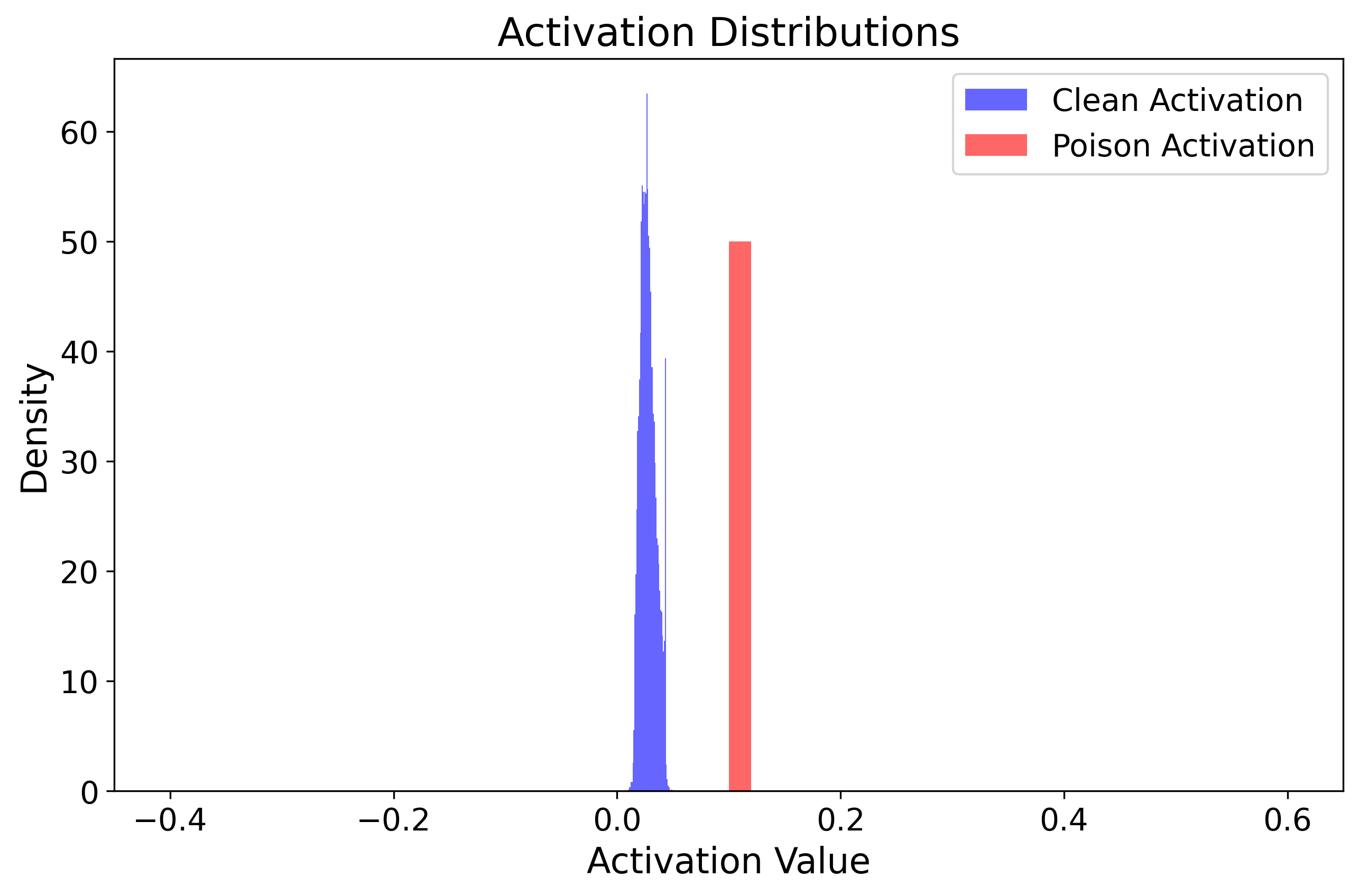}
    \caption{The activation value distributions of clean and backdoor samples 
    }
    \label{fig:nonzero-threshold}
\end{figure}

\begin{figure}[tbp]
    \centering
    \includegraphics[width=0.235\textwidth]{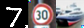}
    \includegraphics[width=0.235\textwidth]{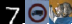}
    \caption{Backdoor triggers on the three benchmark datasets. The triggers are the squared regions on each corner of images. The first three images show trigger samples generated by the trigger optimization attack, while remaining images displays trigger samples generated by the weight optimization attack.}
    \label{fig:trigger-demo}
\end{figure}




\section{Additional Results}
\label{sec:nonzero-threshold}
In \Cref{fig:nonzero-threshold}, we show the distribution of activation values in clean and backdoor samples for the CIFAR10 ResNet model. We can clearly see a nonzero threshold that separates the clean and backdoor activations, and such a threshold can be obtained by slightly tuning the scaling factor $s$ in Eq.~\eqref{eq:single_opt_sol}.

In \Cref{fig:trigger-demo}, we show the sample backdoored images for the MNIST, GTSRB and CIFAR10 datasets. 

\revision{
\shortsection{Fine-Tuning with NAD and DT}
Beyond defenses covered in \Cref{sec:evade_defense}, we also evaluated Neural Attention Distillation (NAD) \cite{li2021neural} and Democratic Training (DT) \cite{sun2025democratic}. NAD is a defense previously shown to underperform compared to our evaluated baseline BEAGLE \cite{cheng2023beagle}. DT, while originally designed for universal adversarial perturbations, was assessed for its potential applicability against backdoor attacks. 
The results, presented in \Cref{tab:ft-fp-results-reviewera}, demonstrate that both defenses have almost no impact on the attack effectiveness. Specifically, two variants of the 1-N attack retained 100\% effectiveness against both NAD and DT. The more generalized 3-N attack saw only a slight drop in effectiveness (around 1\%) for GTSRB and CIFAR10. On ImageNet, DT performed slightly better, reducing the ASR from 97\% before the defense to 90\% afterward, yet this came at the cost of a 6\% drop in clean accuracy. More importantly, an ASR of 90\% still indicates effectiveness of the defense in practice.
}

\begin{table}[htbp]
\centering
\caption{Effectiveness of HAMLOCK against Neural Attention Distillation and Democratic Training on different datasets. The model architecture is ResNet-18. ``NDA'' denotes Neural Attention Distillation, ``DT'' denotes Democratic Training. 
}
\scalebox{0.8}{

\begin{tabular}{ccccccc}
    \toprule
    Datasets & Attack
    & ASR (\%)
    & \multicolumn{2}{c}{Clean Acc (\%)}
    & \multicolumn{2}{c}{ASR (\%)} \\
    \cmidrule(lr){3-3}
    \cmidrule(lr){4-5}
    \cmidrule(lr){6-7}
     & 
     & w/o Defense
     & NDA & DT
     & NDA & DT \\
    \midrule

    \multirow{3}{*}{GTSRB}
    & 1N-trigger & 100.0 & 93.5 & 89.4 & 100.0 & 100.0 \\
    & 1N-weight  & 100.0 & 93.5 & 89.4 & 100.0 & 100.0 \\
    & 3N-weight  & 97.0 & 96.1 & 92.2 & 95.7 & 95.2 \\

    \midrule
    \multirow{3}{*}{CIFAR10}
    & 1N-trigger & 100.0 & 91.3 & 89.9 & 100.0 & 100.0 \\
    & 1N-weight  & 100.0 & 91.4 & 90.0 & 100.0 & 100.0 \\
    & 3N-weight  & 96.0 & 90.8 & 90.7 & 95.4 & 94.8 \\

    \midrule
    \multirow{3}{*}{Imagenet}
    & 1N-trigger & 100.0 & 65.0 & 58.6 & 100.0 & 100.0 \\
    & 1N-weight  & 100.0 & 65.2 & 59.2 & 100.0 & 100.0 \\
    & 3N-weight  & 97.0 & 65.0 & 59.5 & 95.4 & 90.1 \\

    \bottomrule
\end{tabular}

}
\label{tab:ft-fp-results-reviewera}
\end{table}

\revision{
\section{Algorithm}\label{sec:appendix-alg}
Algorithm \ref{alg:multi_neuron_hamlock} outlines the HAMLOCK multi‑neuron weight optimization process described in \Cref{sec:multi-neuron}.

\begin{algorithm}[!h]
\caption{Multi-Neuron Weight Optimization}
\label{alg:multi_neuron_hamlock}

    \begin{algorithmic}[1]
    \State \textbf{Input:} Pre-trained Model $f_{\theta}$, Trigger $\delta$, Scaling Factor $s$, Number of Layers $r$, Neurons per Layer $k$, Calibration Set $\mathcal{D}_{cal}$ ($|D_{cal}|$ = $M$), Accuracy Drop Tolerance $\tau$
    \State \textbf{Output:} Backdoored Model $f_{\theta'}$

    \State $x' \gets x \cdot (1-m) + m \cdot \delta$
    \State Randomly select $r$ layers from $f_{\theta}$. \Comment{Layer Section}
    
    \For{each selected layer $l$}
        \State $\mathcal{C} \gets \emptyset$
        
        \For{each neuron $j$ in layer $l$}
            \State Temporarily set $a_j(x) \gets 0$ for all $x \in \mathcal{D}_{cal}$
            \State Measure accuracy drop $\Delta_{\text{acc}}$
            \If{$\Delta_{\text{acc}} < \tau$}
                \State $\mu_{clean} \gets \frac{1}{M}\sum a_j(x)$
                \State $\mu_{bd} \gets \frac{1}{M}\sum a_j(x')$

                \State $S_j \gets |\mu_{bd} - \mu_{clean}|$
                \State $\mathcal{C} \gets \mathcal{C} \cup \{(j, S_j)\}$
            \EndIf
        \EndFor
        
        \State $\mathcal{T} \gets \operatorname{TopK}(\mathcal{C}, k)$ \Comment{Candidate Selection}
        
        \For{each target neuron $j \in \mathcal{T}$}
        \State $w_j$: weights connecting layer $l-1$ to neuron $j$.
        \For{each connection $i$ from layer $l-1$}
            \State $w'_{ji} \gets s \cdot |w_{ji}| \cdot \text{sgn}( \bar{a_i}(x') - \bar{a_i}(x))$ 
        \EndFor
        \State Update neuron $j$ weights in $f_{\theta}$ with $W'_j$.
    \EndFor
    \EndFor
    
    \State \textbf{return} $f_{\theta}$
    \end{algorithmic}
\end{algorithm}
}

\end{document}